\documentclass[twocolumn]{aastex63}

\received{25 Nov 2020}
\revised{19 Feb 2021}
\accepted{01 Mar 2021}
\submitjournal{AJ}

\shorttitle{A transiting warm giant planet around the young active star TOI-201}
\shortauthors{Hobson et al.}
\graphicspath{{./}{figures/}}

\begin{document}

\title{A transiting warm giant planet around the young active star TOI-201}

\correspondingauthor{Melissa J. Hobson}
\email{mhobson@astro.puc.cl}

\author[0000-0002-5945-7975]{Melissa J. Hobson}
\affiliation{Millennium Institute for Astrophysics, Chile}
\affiliation{Instituto de Astrofísica, Facultad de Física, Pontificia Universidad Católica de Chile, Av. Vicuña Mackenna 4860, 782-0436 Macul, Santiago, Chile}

\author{Rafael Brahm}
\affiliation{Millennium Institute for Astrophysics, Chile}
\affiliation{Facultad de Ingeniería y Ciencias, Universidad Adolfo Ibáñez, Av. Diagonal las Torres 2640, Peñalolén, Santiago, Chile}

\author{Andrés Jordán}
\affiliation{Millennium Institute for Astrophysics, Chile}
\affiliation{Facultad de Ingeniería y Ciencias, Universidad Adolfo Ibáñez, Av. Diagonal las Torres 2640, Peñalolén, Santiago, Chile}

\author{Nestor Espinoza}
\affiliation{Space Telescope Science Institute, 3700 San Martin Drive, Baltimore, MD 21218, USA}

\author{Diana Kossakowski}
\affiliation{Max-Planck-Institut für Astronomie, Königstuhl 17, 69117 Heidelberg, Germany}

\author[0000-0002-1493-300X]{Thomas Henning}
\affiliation{Max-Planck-Institut für Astronomie, Königstuhl 17, 69117 Heidelberg, Germany}

\author{Felipe Rojas}
\affiliation{Millennium Institute for Astrophysics, Chile}
\affiliation{Instituto de Astrofísica, Facultad de Física, Pontificia Universidad Católica de Chile, Av. Vicuña Mackenna 4860, 782-0436 Macul, Santiago, Chile}

\author[0000-0001-8355-2107]{Martin Schlecker}
\affiliation{Max-Planck-Institut für Astronomie, Königstuhl 17, 69117 Heidelberg, Germany}

\author{Paula Sarkis}
\affiliation{Max-Planck-Institut für Astronomie, Königstuhl 17, 69117 Heidelberg, Germany}

\author{Trifon Trifonov}
\affiliation{Max-Planck-Institut für Astronomie, Königstuhl 17, 69117 Heidelberg, Germany}

\author[0000-0002-5113-8558]{Daniel Thorngren}
\affiliation{University of Montréal, Montréal, QC, Canada}

\author[0000-0002-9319-3838]{Avraham Binnenfeld}
\affiliation{Porter School of the Environment and Earth Sciences, Raymond and Beverly Sackler Faculty of Exact Sciences, Tel Aviv University,Tel Aviv, 6997801, Israel}

\author[0000-0001-9298-8068]{Sahar Shahaf}
\affiliation{School  of  Physics  and  Astronomy,  Raymond  and  Beverly  Sackler  Faculty  of  Exact  Sciences,  Tel  Aviv  University,  Tel  Aviv,6997801, Israel}

\author[0000-0003-3173-3138]{Shay Zucker}
\affiliation{Porter School of the Environment and Earth Sciences, Raymond and Beverly Sackler Faculty of Exact Sciences, Tel Aviv University,Tel Aviv, 6997801, Israel}

\author[0000-0003-2058-6662]{George~R.~Ricker}
\affiliation{Department of Physics and Kavli Institute for Astrophysics and Space Research, Massachusetts Institute of Technology, Cambridge, MA 02139, USA}


\author[0000-0001-9911-7388]{David~W.~Latham}
\affiliation{Center for Astrophysics \textbar \ Harvard \& Smithsonian, 60 Garden Street, Cambridge, MA 02138, USA}

\author[0000-0002-6892-6948]{S.~Seager}
\affiliation{Department of Physics and Kavli Institute for Astrophysics and Space Research, Massachusetts Institute of Technology, Cambridge, MA 02139, USA}
\affiliation{Department of Earth, Atmospheric and Planetary Sciences, Massachusetts Institute of Technology, Cambridge, MA 02139, USA}
\affiliation{Department of Aeronautics and Astronautics, MIT, 77 Massachusetts Avenue, Cambridge, MA 02139, USA}

\author[0000-0002-4265-047X]{Joshua~N.~Winn}
\affiliation{Department of Astrophysical Sciences, Princeton University, 4 Ivy Lane, Princeton, NJ 08544, USA}

\author[0000-0002-4715-9460]{Jon~M.~Jenkins}
\affiliation{NASA Ames Research Center, Moffett Field, CA 94035, USA}


\author[0000-0003-3216-0626]{Brett Addison}
\affiliation{University of Southern Queensland, Centre for Astrophysics, West Street, Toowoomba, QLD 4350 Australia}



\author{François Bouchy}
\affiliation{Geneva Observatory, University of Geneva, Chemin des Mailettes 51, 1290 Versoix, Switzerland}

\author{Brendan P. Bowler}
\affiliation{Department of Astronomy, The University of Texas at Austin, TX 78712, USA}


\author[0000-0002-7611-8772]{Joshua T. Briegal}
\affiliation{Astrophysics Group, Cavendish Laboratory, J.J. Thomson Avenue, Cambridge CB3 0HE, UK}

\author[0000-0001-7904-4441]{Edward M. Bryant} 
\affiliation{Department of Physics, University of Warwick, Gibbet Hill Road, Coventry CV4 7AL, UK}
\affiliation{Centre for Exoplanets and Habitability, University of Warwick, Gibbet Hill Road, Coventry CV4 7AL, UK}

\author[0000-0001-6588-9574]{Karen A.\ Collins}
\affiliation{Center for Astrophysics \textbar \ Harvard \& Smithsonian, 60 Garden Street, Cambridge, MA 02138, USA}

\author[0000-0002-6939-9211]{Tansu~Daylan}
\affiliation{Department of Physics and Kavli Institute for Astrophysics and Space Research, Massachusetts Institute of Technology, Cambridge, MA 02139, USA}
\affiliation{Kavli Fellow}

\author{Nolan Grieves}
\affiliation{Geneva Observatory, University of Geneva, Chemin des Mailettes 51, 1290 Versoix, Switzerland}

\author[0000-0002-1160-7970]{Jonathan Horner}
\affiliation{University of Southern Queensland, Centre for Astrophysics, West Street, Toowoomba, QLD 4350 Australia}

\author{Chelsea Huang}
\affiliation{Department of Physics and Kavli Institute for Astrophysics and Space Research, Massachusetts Institute of Technology, Cambridge, MA 02139, USA}

\author[0000-0002-7084-0529]{Stephen R. Kane}
\affiliation{Department of Earth and Planetary Sciences, University of California, Riverside, CA 92521, USA}

\author[0000-0003-0497-2651]{John Kielkopf}
\affiliation{Department
 of Physics and Astronomy, University of Louisville, Louisville, KY 40292, USA}

\author{Brian McLean}
\affiliation{Space Telescope Science Institute, Mikulski Archive for Space Telescopes, Baltimore, MD, 21218, USA}

\author[0000-0002-7830-6822]{Matthew W. Mengel}
\affiliation{University of Southern Queensland, Centre for Astrophysics, West Street, Toowoomba, QLD 4350 Australia}

\author[0000-0002-5254-2499]{Louise D. Nielsen}
\affiliation{Geneva Observatory, University of Geneva, Chemin des Mailettes 51, 1290 Versoix, Switzerland}

\author[0000-0002-4876-8540]{Jack Okumura}
\affiliation{University of Southern Queensland, Centre for Astrophysics, West Street, Toowoomba, QLD 4350 Australia}

\author{Peter Plavchan}
\affiliation{George Mason
 University, 4400 University Drive MS 3F3, Fairfax, VA 22030, USA}

\author[0000-0002-1836-3120]{Avi Shporer}
\affiliation{Department
 of Physics and Kavli Institute for Astrophysics and Space Research, Massachusetts Institute of Technology, Cambridge, MA 02139, USA}

\author[0000-0002-2386-4341]{Alexis M. S. Smith}
\affiliation{Institute of Planetary Research, German Aerospace Center, Rutherfordstrasse 2, D-12489, Berlin, Germany}

\author{Rosanna Tilbrook}
\affiliation{School of Physics and Astronomy, University of Leicester, University Road, Leicester LE1 7RH, UK}

\author[0000-0002-7595-0970]{C.G. Tinney}
\affiliation{Exoplanetary
 Science at UNSW, School of Physics, UNSW Sydney, NSW 2052, Australia}

\author[0000-0002-6778-7552]{Joseph D. Twicken}
\affiliation{SETI Institute, Mountain View, CA  94043, USA}
\affiliation{NASA Ames Research Center, Moffett Field, CA  94035, USA}

\author{Stéphane Udry}
\affiliation{Geneva Observatory, University of Geneva, Chemin des Mailettes 51, 1290 Versoix, Switzerland}

\author[0000-0003-3993-7127]{Nicolas Unger}
\affiliation{Geneva Observatory, University of Geneva, Chemin des Mailettes 51, 1290 Versoix, Switzerland}

\author{Richard West}
\affiliation{Department of Physics, University of Warwick, Gibbet Hill Road, Coventry CV4 7AL, UK}

\author[0000-0001-9957-9304]{Robert
 A. Wittenmyer}
\affiliation{University of Southern Queensland, Centre for Astrophysics, West Street, Toowoomba, QLD 4350 Australia}

\author[0000-0002-5402-9613]{Bill Wohler}
\affiliation{SETI Institute, Mountain View, CA  94043, USA}
\affiliation{NASA Ames Research Center, Moffett Field, CA  94035, USA}

\author[0000-0001-7294-5386]{Duncan J. Wright}
\affiliation{University of Southern Queensland, Centre for Astrophysics, West Street, Toowoomba, QLD 4350 Australia}



\begin{abstract}

We present the confirmation of the eccentric warm giant planet TOI-201 b, first identified as a candidate in \textit{TESS} photometry (Sectors 1-8, 10-13, and 27-28) and confirmed using ground-based photometry from NGTS and radial velocities from FEROS, HARPS, CORALIE, and \textsc{Minerva}-Australis. TOI-201 b orbits a young ($\mathrm{0.87^{+0.46}_{-0.49} \, Gyr}$) and bright(V=9.07 mag) F-type star with a $\mathrm{52.9781 \, d}$ period. The planet has a mass of $\mathrm{0.42^{+0.05}_{-0.03}\, M_J}$, a radius of $\mathrm{1.008^{+0.012}_{-0.015}\, R_J}$, and an orbital eccentricity of $0.28^{+0.06}_{-0.09}$; it appears to still be undergoing fairly rapid cooling, as expected given the youth of the host star. The star also shows long-term variability in both the radial velocities and several activity indicators, which we attribute to stellar activity. The discovery and characterization of warm giant planets such as TOI-201 b is important for constraining formation and evolution theories for giant planets.

\end{abstract}



\section{Introduction} \label{sec:intro}
Transiting warm giants - that is, planets with $\mathrm{R_P \gtrsim 0.8\, R_J}$ and $\mathrm{10\, d \lesssim P \lesssim 100 \, d}$ - are particularly important for understanding the formation and evolution of giant planets. Unlike hot Jupiters - i.e., planets with $\mathrm{R_P \gtrsim 0.8\, R_J}$ and $\mathrm{P \lesssim 10 \, d}$ - which are inflated by mechanisms that are still unclear but likely connected to irradiation \citep[e.g.][]{Sarkis20} these more distant planets are less strongly irradiated by their host star, meaning their size and mass can be effectively modelled by their metallicity \citep[e.g.][]{Thorngren19}. 
Both hot and warm Jupiters are unlikely to form in situ, but rather are expected to have formed in the outer regions of the disk and migrated to their current locations; the main mechanisms proposed are gas disk migration and high eccentricity migration (see \citealt{Dawson18} for a comprehensive review). However, hot Jupiters' orbital histories are affected by tidal evolution, which can erase traces of past interactions between planets; this is not the case for warm Jupiters. Therefore, this population of planets preserves valuable information for the study of giant planet formation in their physical and orbital parameters. These parameters can be characterized through the combination of photometry and radial velocities. 

While the current sample of warm giants is still small, the \textit{Transiting Exoplanet Survey Satellite} \citep[\textit{TESS},][]{Ricker15} is expected to detect hundreds of such planets \citep{Sullivan15, Barclay18}. The Warm gIaNts with tEss collaboration \citep[WINE, e.g.][]{Brahm19PARSEC, Jordan20} is using a network of photometric and spectroscopic facilities to follow up, confirm, and characterize warm giant candidates from \textit{TESS}. One such candidate is TOI-201.01.

In this paper, we present the confirmation and characterization of the warm giant TOI-201 b, orbiting the young F-type star TOI-201. The paper is organized as follows. We present the observational data in Section \ref{sec:obs}. In Section \ref{sec:analysis} we analyse the data, and characterize the star and planet. Finally, our results are discussed in Section \ref{sec:disc}.

\section{Observations} \label{sec:obs}

\subsection{Photometric data}

\subsubsection{\textit{TESS}}

TOI-201 was observed by the \textit{TESS} mission between 25 July 2018 and 17 July 2019, in Sectors 1, 2, 3, 4, 5, 6, 7, 8, 10, 11, 12, and 13, using camera 4. CCD 1 was used for Sectors 3, 4, and 5; CCD 2 for Sectors 6, 7, and 8; CCD 3 for Sectors 10, 11, and 12; and CCD 4 for Sectors 1, 2, and 13. Currently, it is also being observed as part of the extended mission; the light curves from Sectors 27 and 28, which were observed between 5 July and 25 August 2020, were also incorporated into the analysis. Camera 4 in CCD 4 was used for both these Sectors.

\begin{figure*}[htb!]
    \centering
    \includegraphics[width=1\hsize]{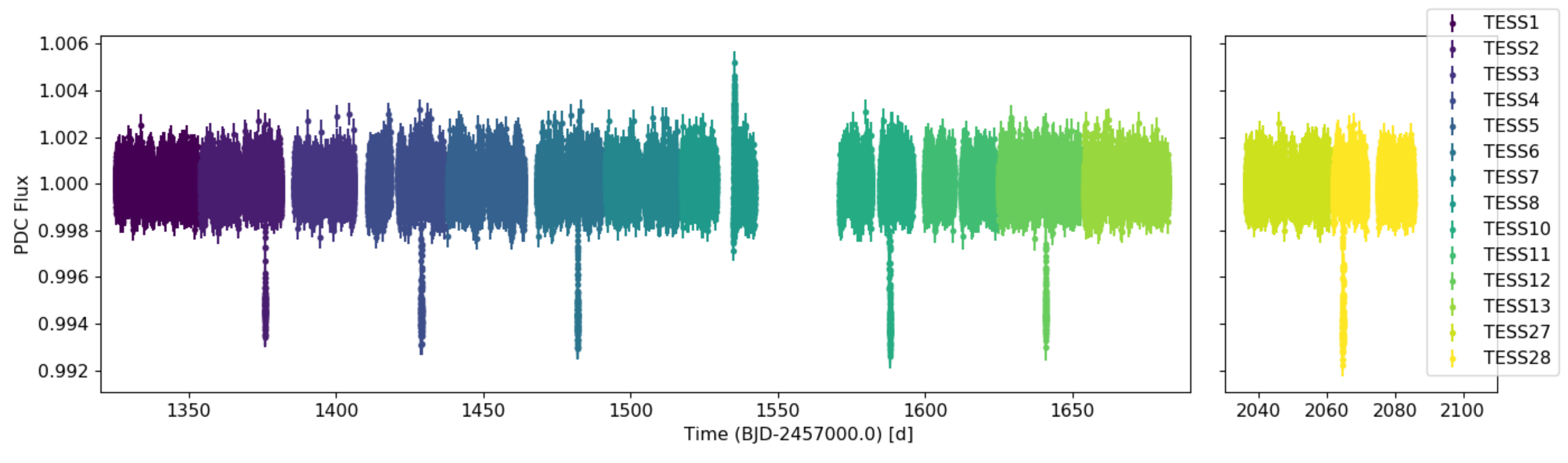}  
    \includegraphics[width=1\hsize]{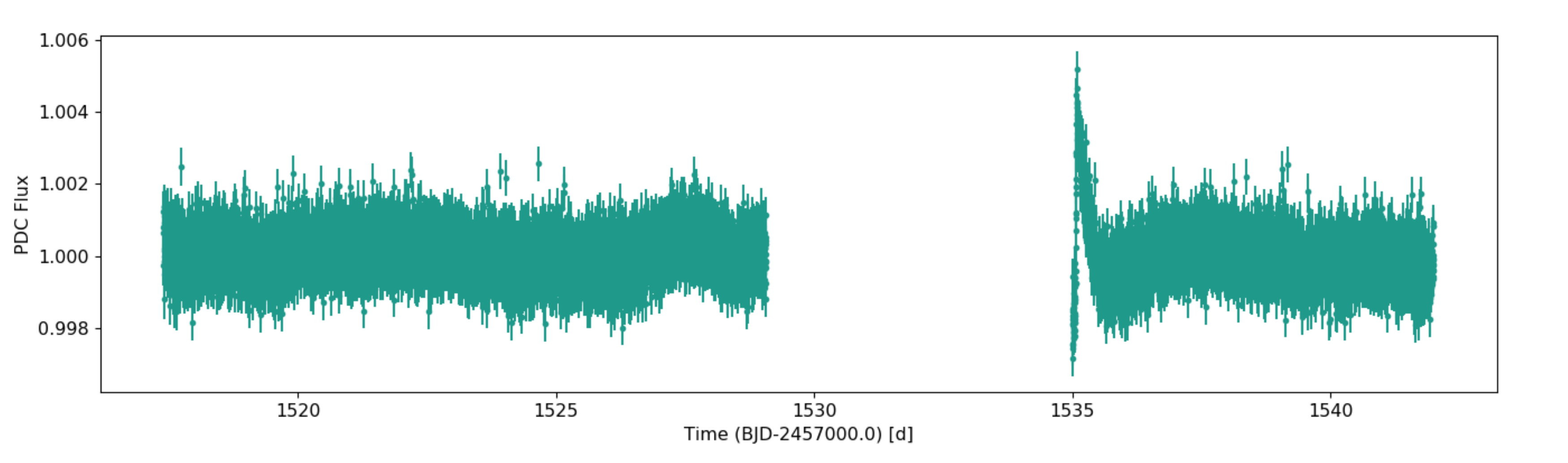}
    \caption{Top: \textit{TESS} light curve for TOI-201, for all Sectors. The data are colour-coded by Sectors, and flagged points have been removed. The left subplot shows the twelve Sectors from the prime mission, and the right subplot the two Sectors from the extended mission. Bottom: Zoom to Sector 8, showing the 5-day gap in the light curve, and the partial transit of TOI-201.01 at the edge of the gap.}
    \label{fig:lightcurve}
\end{figure*}

The 2-minute cadence data for TOI-201 were processed in the TESS Science Processing Operation Center \citep[SPOC,][]{Jenkins16} pipeline. Two potential transit signals were identified in the SPOC transit search \citep{Jenkins02, Jenkins10, Jenkins20} of the TOI-201 light curve. These were designated as TESS Objects of Interest (TOIs) by the TESS Science Office based on SPOC data validation results \citep{Twicken18, Li19} indicating that both signals were consistent with transiting planets. The planetary candidates are listed in the ExoFOP-TESS archive\footnote{Located at \url{https://exofop.ipac.caltech.edu/tess/target.php?id=350618622}}: TOI-201.01, with a period of $\mathrm{P_{01} = 52.978306\, d}$, and TOI-201.02, with a period of $\mathrm{P_{02} = 5.849173\, d}$. The full PDCSAP light curve \citep{Stumpe12, Stumpe14, Smith12}, obtained from the MAST archive, is shown in Fig. 1 (top panel). Any points that were flagged as being of low quality have been removed. Six full transits of TOI-201.01 are clearly visible, in Sectors 2, 4, 6, 10, 12, and 28. A partial transit is also visible in Sector 8, at the edge of a five day gap in the light curve (Fig. \ref{fig:lightcurve}, bottom panel). This gap is due to an instrument turn-off between TJD\footnote{TESS Julian Date, TJD = BJD - 2457000.0} 1531.74 and TJD 1535.00, caused by an interruption in communications between the instrument and spacecraft. 
The flux is clearly overestimated for the first $\mathrm{\sim 0.5 \, d}$ after the gap. This artifact appears to have been introduced in the SPOC PDC processing, as the SAP light curve shows a low flux level for this period, which is likely due to the camera temperature change of $\mathrm{\sim 20^{\circ}}$ during the instrument turn-off, as detailed in the \textit{TESS} Data Release Notes for Sector 8, DR10\footnote{Archived at \url{https://archive.stsci.edu/missions/tess/doc/tess_drn/tess_sector_08_drn10_v02.pdf}}. 
The transits of TOI-201.02 have a much shallower depth (only $\mathrm{0.128 \pm 0.013 \, mmag}$, compared to  $\mathrm{6.843 \pm 0.058 \, mmag}$ for TOI-201.01), and are hence not visually obvious in the light curve.

\subsubsection{NGTS}

\begin{figure}[htb]
    \centering
    \includegraphics[width=1\hsize]{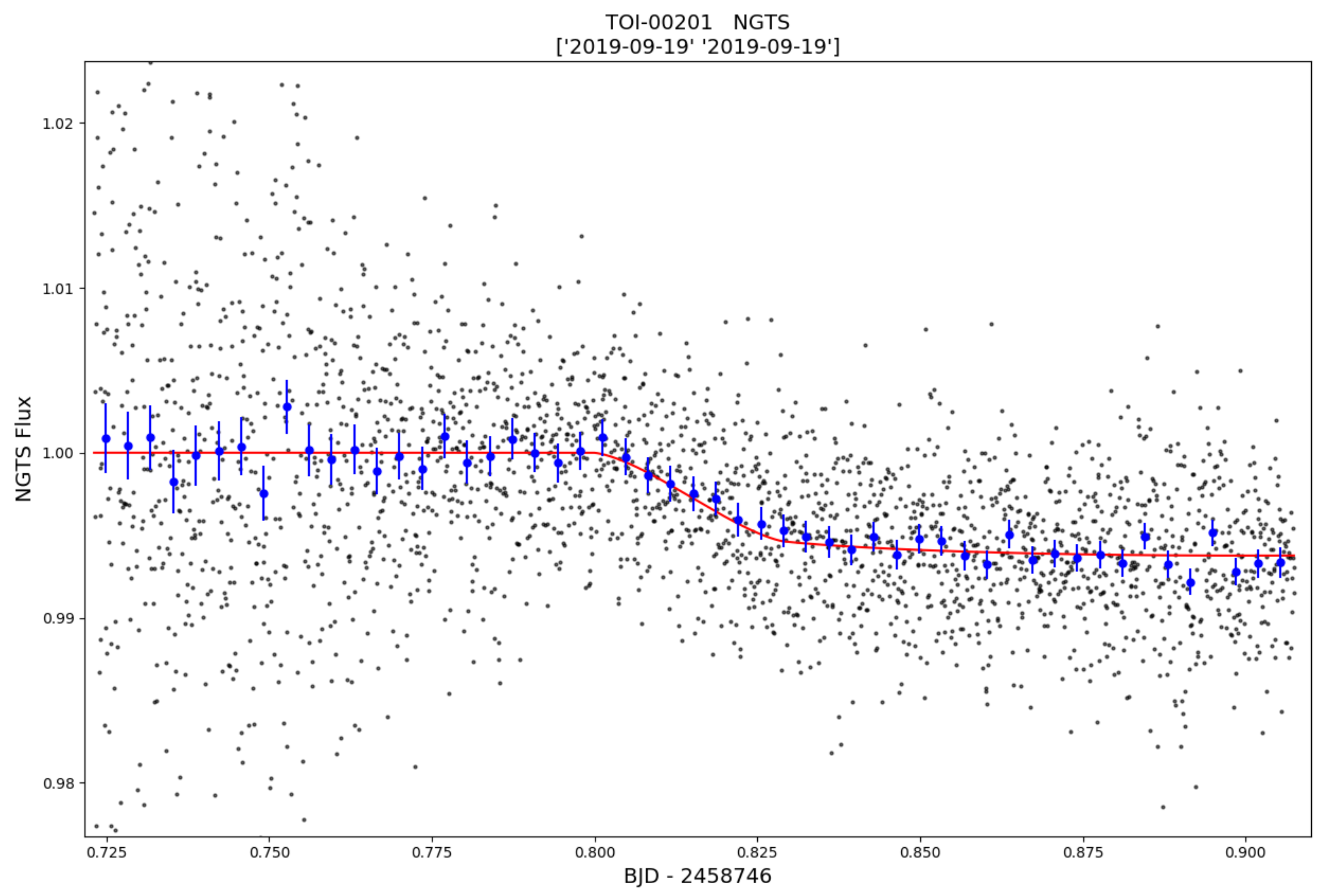}
    \caption{NGTS light curve for TOI-201, observed on 19th September 2019. The blue points show the data binned to 5 minutes. The red line indicates the predicted model from the \textit{TESS} data, using ephemeris of $\mathrm{T_0 = 1376.052045}$, $\mathrm{P = 52.978306 \, d}$.}
    \label{fig:NGTS}
\end{figure}

Due to the relatively large $21\arcsec$ pixel size of \textit{TESS}, nearby companions can contaminate its photometry. It is therefore necessary to identify possible contaminating sources, which can create false positives or dilute the transits. Ground-based photometry plays an important role in disentangling these false positives. In this context, The Next Generation Transit Survey \citep[NGTS,][]{2018:wheatley:ngts} observed TOI-201 on 19th September 2019, obtaining a clear ingress signal (Fig. \ref{fig:NGTS}). Two telescopes were used, with the custom NGTS filter (520-890 nm). A total of 2484 images were obtained with an exposure time of 10 seconds, for an overall cadence (exposure, read out, etc.) of 13 seconds. The images were reduced with a custom aperture photometry pipeline \citep{Bryant20}, using an aperture with a radius of 7 pixels ($35\arcsec$). 

\subsection{High-resolution imaging}

\begin{figure}[htb]
    \centering
    \includegraphics[width=1\hsize]{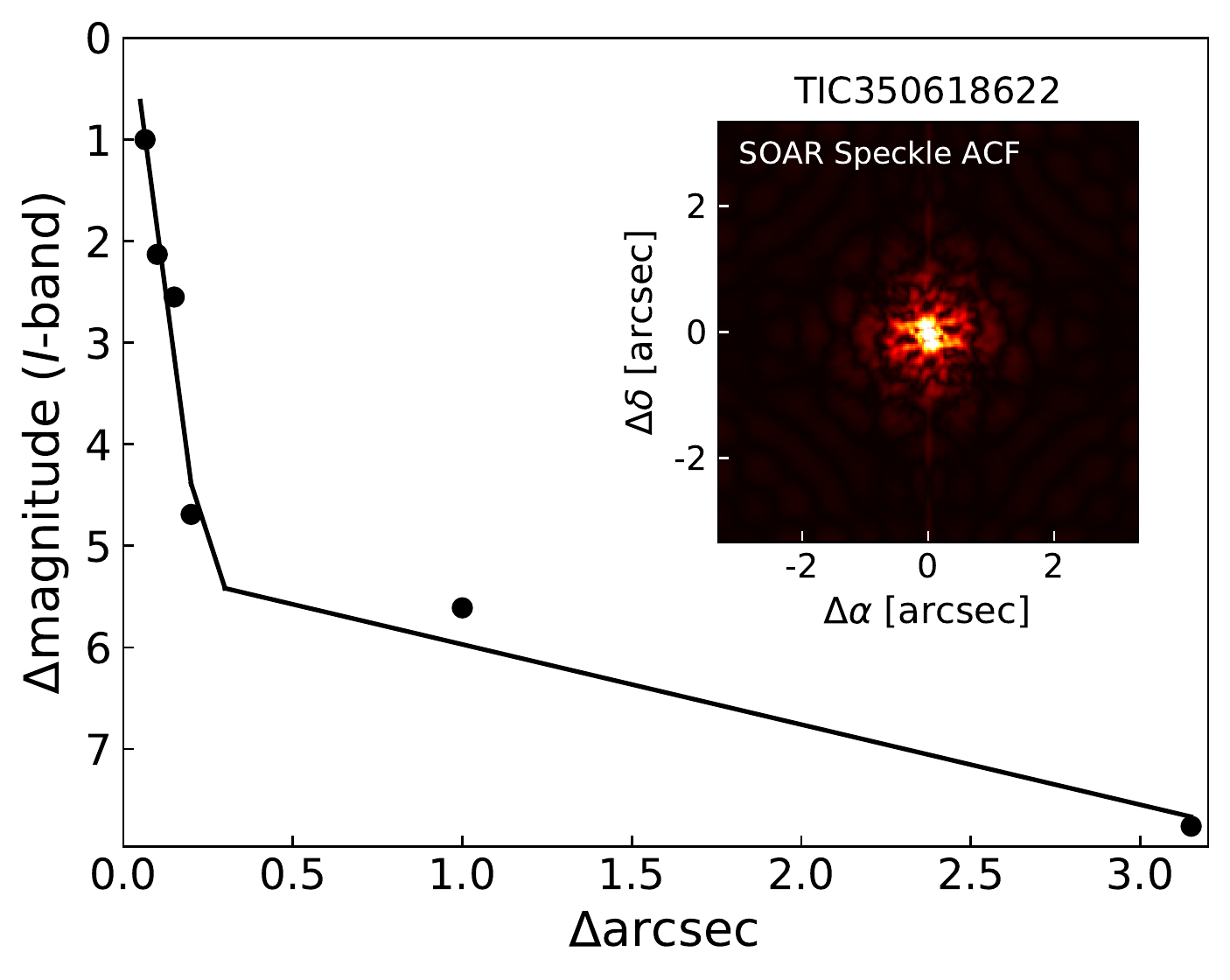}
    \caption{Contrast curve and speckle auto-correlation function from the HRCam at SOAR for TOI-201. The black points and solid line indicate the $5\sigma$ contrast curve; the inset shows the speckle auto-correlation function.}
    \label{fig:speckle-imaging}
\end{figure}

High-resolution imaging is a valuable tool for rejecting false positive scenarios for \textit{TESS} candidates. In this context, the SOAR \textit{TESS} survey \citep{Ziegler20} has observed \textit{TESS} planet candidate hosts with speckle imaging using the high-resolution camera (HRCam) imager on the 4.1-m Southern Astrophysical Research (SOAR) telescope at Cerro Pachón, Chile \citep{Tokovinin18}. TOI-201 was observed on the night of 18 February 2019, with no nearby sources detected within $3\arcsec$. The contrast curve and auto-correlation function are shown in Fig. \ref{fig:speckle-imaging}.

\subsection{Spectroscopic data}

The WINE consortium carried out spectroscopic follow-up of TOI-201 with the FEROS and HARPS spectrographs. In addition to these, we also obtained data from the CORALIE and \textsc{Minerva}-Australis teams. Finally, TOI-201 was observed once with NRES, for the purpose of stellar parameter determination.

\subsubsection{FEROS}

We obtained 52 spectra with the FEROS spectrograph \citep{Kaufer99}, which is mounted on the MPG 2.2m telescope at La Silla Observatory and has a resolving power of $\mathrm{R=50\,000}$, between 26 November 2018 and 8 March 2020. The observations were performed in the Object-Calibration mode, with an exposure time of 5 minutes. The spectra, which have a median signal to noise ratio (SNR) of 165, were processed with the \texttt{CERES} pipeline \citep{Brahm17CERES}. From this pipeline, we obtain both radial velocities (RV) and activity indicators. Specifically, we compute the bisector of the CCF (BIS), which traces photospheric activity \citep[e.g.][]{Queloz01}, and the H$_\alpha$, log($R^\prime_{HK}$), Na~II, and He~I activity indices, which trace chromospheric activity. For H$_\alpha$, we used the definition of \citet{Boisse09}. As TOI-201 is an F-type star, we used the regions defined by \citet{Duncan91} and the calibrations of \citet{Noyes84} for log($R^\prime_{HK}$). For Na~II and He~I we followed \citet{Gomes11}. The radial velocities and activity indices computed from the FEROS data are listed in table \ref{tab:feros-data}; the RVs have a median error of $\mathrm{\sigma_{RV} \approx 9.1\, m/s}$. 

\subsubsection{HARPS}

We obtained 39 spectra with the HARPS spectrograph \citep{Mayor03}, which is mounted on the 3.6m telescope at La Silla Observatory, and has a resolving power of $\mathrm{R=120\,000}$. The spectra were obtained between 11 December 2018 and 21 February 2020, under Program IDs 0101.C-0510(D), 0102.C-0451(B), 0103.C-0442(A), and 0104.C-0413(A). The exposures were taken in simultaneous sky mode, with a 10 minute duration. One observation, obtained on 17 January 2020, has an extremely low SNR of 17 and was therefore removed from the analysis. The remaining 38 spectra have a median SNR of 109. As with the FEROS spectra, we processed the HARPS spectra using the \texttt{CERES} pipeline, obtaining the radial velocities and the same set of activity indicators. The radial velocities and activity indices computed from the HARPS data are listed in table \ref{tab:harps-data}; the RVs have a median error of $\mathrm{\sigma_{RV} \approx 2.0\, m/s}$.

\subsubsection{CORALIE}

A total of 10 spectra of TOI-201 were obtained with the CORALIE high resolution spectrograph on the Swiss 1.2 m Euler telescope at La Silla Observatories, Chile \citep{CORALIE}, between 12 November 2018 and 27 March 2019. Each exposure had a duration of 20 minutes yielding a SNR of $\sim 45$. CORALIE is a fibre-fed echelle spectrograph with a $2\arcsec$ science fibre and a secondary fibre with a Fabry-Perot for simultaneous wavelength calibration. We extract radial velocity measurements by cross-correlation of the spectra with a binary G2V template \citep{1996A&AS..119..373B,2002Msngr.110....9P}. We obtain final RV uncertainties for the CORALIE epochs of $\mathrm{\sigma_{RV} \approx 10-15\, m/s}$. BIS, FWHM and other line-profile diagnostics were computed as well using the standard CORALIE DRS. We also compute the $\mathrm{H_\alpha}$ index for each spectrum to check for possible variation in stellar activity. We investigate various scenarios where a blended eclipsing stellar binary can mimic a transiting planet using the line profile diagnostics and find no evidence of such.

\subsubsection{\textsc{Minerva}-Australis}

\textsc{Minerva}-Australis is an array of four PlaneWave CDK700 telescopes located in Queensland, Australia, fully dedicated to the precise radial-velocity follow-up of \textit{TESS} candidates \citep[e.g.][]{toi257,AUMicBrett,toi677}. The four telescopes can be simultaneously fiber-fed to a single KiwiSpec R4-100 high-resolution ($\mathrm{R=80\,000}$) spectrograph \citep{barnes12,addison19}.  
We obtained 62 spectra of TOI-201 in the early days of Minerva-Australis, with a single telescope, between 2 January 2019 and 15 April 2019.  Exposure times were 30 minutes, and on some nights, two consecutive exposures were obtained.  The resulting radial velocities are given in Table ~\ref{tab:minerva-data}. 
Radial velocities for the observations are derived for each telescope by cross-correlation, where the template being matched is the mean spectrum of each telescope. The instrumental variations are corrected by using simultaneous Thorium-Argon arc lamp observations.

\subsubsection{NRES}\label{ssec:NRES}

The Network of Robotic Echelle Spectrographs \citep[NRES,][]{Siverd18} of the Las Cumbres Observatory  \citep[LCOGT, ][]{Brown13} consists of four identical fibre-fed optical echelle spectrographs, with a resolving power of $\mathrm{R \approx 53\,000}$, mounted on globally distributed 1 m telescopes. TOI-201 was observed by NRES at CTIO on 8 December 2018. A SpecMatch analysis \citep{Yee17} was performed on the spectrum, yielding the following stellar parameters: $\mathrm{T_{eff} = 6280 \pm 100 \, K}$, $\mathrm{\log{g} = 4.3 \pm 0.1}$, $\mathrm{Fe/H = 0.21 \pm 0.06}$, $\mathrm{v \sin{i} = 8.4 \pm 2.1 \, kms^{-1}}$, $\mathrm{M_\star = 1.285 \pm 0.064 \, M_\odot}$, and $\mathrm{R_\star = 1.35 \pm 0.17 \, R_\odot}$.

\section{Analysis} \label{sec:analysis}

\subsection{Stellar parameters}\label{sec:starparam}

In order to characterise the host star, we first used the co-added HARPS spectra to determine the atmospheric parameters. We employed the \texttt{ZASPE} code \citep{Brahm17ZASPE}, which compares the observed spectrum to a grid of synthetic models generated from the ATLAS9 model atmospheres \citep{Castelli03} in order to determine the effective temperature $\mathrm{T_{eff}}$, surface gravity $\mathrm{\log{g}}$, metallicity [Fe/H], and projected rotational velocity $\mathrm{v \sin{i}}$. 

Next, we followed the second procedure described in \cite{Brahm19PARSEC} to determine the physical parameters. To summarize, we compare the broadband photometric measurements, converted to absolute magnitudes using the \textit{Gaia} DR2 \citep{GAIA2016, GAIA2018} parallax, with the stellar evolutionary models of \cite{Bressan12}. We use the \texttt{emcee} package \citep{Foreman-Mackey13} to sample the posterior distributions. Through this procedure, we determine the age, mass, radius, luminosity, density, and extinction. We also obtain new values for $\mathrm{T_{eff}}$ and $\mathrm{\log{g}}$, the latter of which is more precise than the one previously determined through \texttt{ZASPE}. Therefore, we iterate the entire procedure, using the new $\mathrm{\log{g}}$ as an additional input parameter for \texttt{ZASPE}. One iteration was sufficient for the $\mathrm{\log{g}}$ value to converge. The stellar parameters and observational properties of TOI-201 are presented in Table \ref{tab:starparams}. The quoted error bars for the stellar parameters are internal errors, computed as the $\pm 1 \sigma$ interval around the median of the posterior for each parameter; they do not take into account any systematic errors on the stellar models. The parameters we derive are consistent at one sigma with those obtained from the NRES spectrum (Section \ref{ssec:NRES}). We find that TOI-201 is a young, F-type star.

\begin{table}[htb]
\begin{center}
\caption{Stellar parameters of TOI-201.}
\label{tab:starparams}
\centering
\begin{tabular}{lcr}
\hline \hline
Parameter & Value                        & Reference \\
\hline
Names     & HD 39474                     &           \\
          & TIC 350618622, TOI-201       & \textit{TESS}      \\
          & J05493641-5454386      & 2MASS     \\
          & 4767547667180525696 & \textit{Gaia} DR2  \\
RA \dotfill (J2000) & $05^h 49^m 36.4138946584^s$  & \textit{Gaia} DR2  \\
DEC \dotfill (J2000) & $-54^h 54^m 38.552783926^s$  & \textit{Gaia} DR2  \\
pm$^{\rm RA}$ \hfill [mas yr$^{-1}$] & 7.731 $\pm$ 0.052 & \textit{Gaia} DR2  \\
pm$^{\rm DEC}$ \hfill [mas yr$^{-1}$] & 66.448 $\pm$ 0.058 & \textit{Gaia} DR2  \\
$\pi$ \dotfill [mas] & 8.7566 $\pm$ 0.0265 & \textit{Gaia} DR2  \\
\hline
T \dotfill [mag] & 8.5822$\pm$ 0.006 & \textit{TESS}      \\
B \dotfill [mag] & 10.104 $\pm$ 0.055 & APASS     \\
V \dotfill [mag] & 9.715 $\pm$ 0.079 & APASS     \\
J \dotfill [mag] & 8.103 $\pm$ 0.029 & 2MASS     \\
H \dotfill [mag] & 7.923 $\pm$ 0.036 & 2MASS     \\
K$_s$ \dotfill [mag] & 7.846 $\pm$ 0.024 & 2MASS     \\
\hline
$T_{\rm eff}$ \dotfill [K] & 6394 $\pm$ 75  & this work \\
Spectral type  \dotfill & F6V  & PM13 \\
Fe/H \dotfill [dex] & 0.240 $\pm$ 0.036 & this work \\
$\log{g}$ \dotfill [dex] & 4.318 $\pm$ 0.014 & this work \\
$v \sin{i}$ \dotfill [$\mathrm{kms^{-1}}$] & 9.52 $\pm$ 0.278 & this work \\
R$_\star$ \dotfill [R$_\odot$] & 1.317 $\pm$ 0.011 & this work \\
M$_\star$ \dotfill [M$_\odot$] & 1.316 $\pm$ 0.027 & this work \\
L$_\star$ \dotfill [L$_\odot$] & 2.6 $\pm$ 0.1 & this work \\
$\rho_\star$ \dotfill [g cm$^{-3}$] & 0.81 $\pm$ 0.03 & this work \\
Age \dotfill [Gyr] & $0.87^{+0.46}_{-0.49}$ & this work \\
A$_V$ \dotfill [mag] & 0.11 $\pm$ 0.05 & this work \\
$\log R'_{\rm hk}$ \dotfill & -4.76 $\pm$ 0.04 & this work \\
\hline
\end{tabular}
\end{center}
    \textit{TESS}: \textit{TESS} Input Catalog \citep{Stassun2019}; 2MASS: Two-micron All Sky Survey \citep{2MASS}; \textit{Gaia} DR2: \textit{\textit{Gaia}} Data Release 2 \citep{GAIA2016, GAIA2018}; APASS: AAVSO Photometric All-Sky Survey \citep{APASS}; PM13: using the tables of \cite{Pecaut13}.
\end{table}

\subsection{Radial velocity analysis}\label{sec:rvonly}

In this section, we present the analysis of the FEROS and HARPS spectra, all of which were processed using the \texttt{CERES} pipeline. For homogeneity, we do not include the CORALIE and \textsc{Minerva}-Australis data in this section, as they were processed using the respective instrument pipelines. Additionally, the FEROS and HARPS data have longer temporal baselines, covering both the 2018-2019 and 2019-2020 observing seasons, whereas CORALIE and \textsc{Minerva}-Australis only observed TOI-201 during the 2018-2019 season.

\begin{figure}
    \centering
    \includegraphics[width=1\hsize]{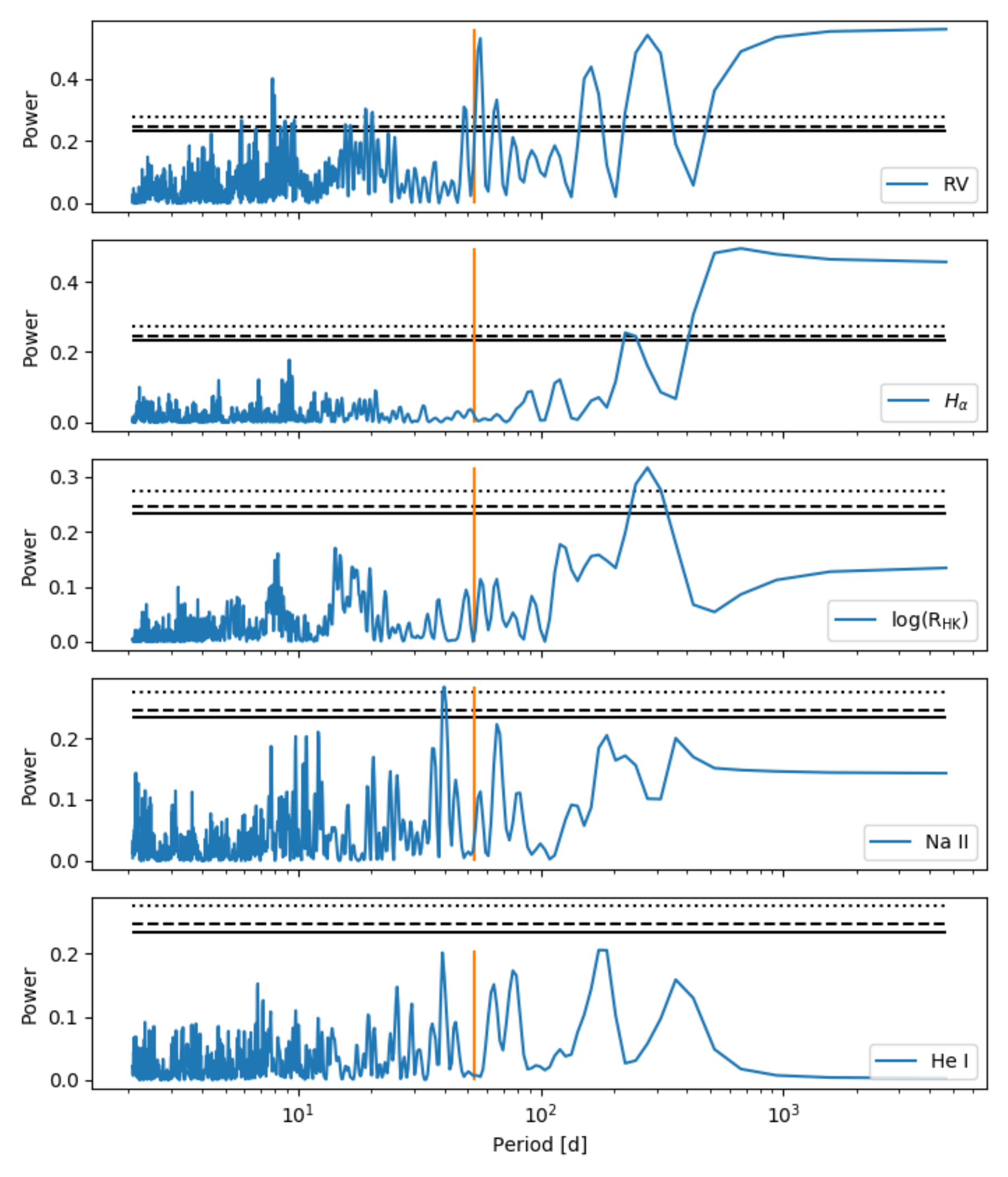}
    \caption{GLS Periodograms of the joint FEROS-HARPS time series for (top to bottom) radial velocities, H$_\alpha$, log($R^\prime_{HK}$), Na~II, and He~I. The solid, dashed, and dotted horizontal black lines indicate the 1\%, 0.5\% and 0.01\% FAP levels respectively. The vertical orange line indicates the period of the planetary candidate TOI-201.01.}
    \label{fig:periodograms-all}
\end{figure}

The GLS periodograms of the joint FEROS and HARPS RV and activity indices time series are presented in Figure \ref{fig:periodograms-all}. The $\mathrm{52.9\, d}$ period of the planetary candidate TOI-201.01 is highlighted. There is a clear, highly significant peak close to this period in the RV periodogram, as well as several significant long-period peaks. We do not find significant peaks close to $\mathrm{\sim 53\, d}$ for any of the activity indices, indicating that the signal in the RVs is likely to be planetary in nature. On the other hand, there are several short- and long-term signatures in the activity indices. There is no clear rotation period in the light curve according to \cite{CantoMartins20}, but given the radius and $v \sin{i}$ reported in Table \ref{tab:starparams}, we may expect a period of $\mathrm{\sim 7 \,d}$ (assuming no misalignment); there is a peak at $\mathrm{\sim 8 \,d}$ in the RVs, and hints of peaks at the same period in the log($R^\prime_{HK}$) Na~II and He~I periodograms. There are also long-period peaks at $\mathrm{\sim 300 \,d}$ in both the H$_\alpha$ and log($R^\prime_{HK}$) periodograms, similar to the one seen in the RVs. Additionally, there are peaks at $\mathrm{\sim 40 \,d}$ in the Na~II and He~I periodograms, with no obvious correspondence in the RVs.  

\begin{figure}
    \centering
    \includegraphics[width=1\hsize]{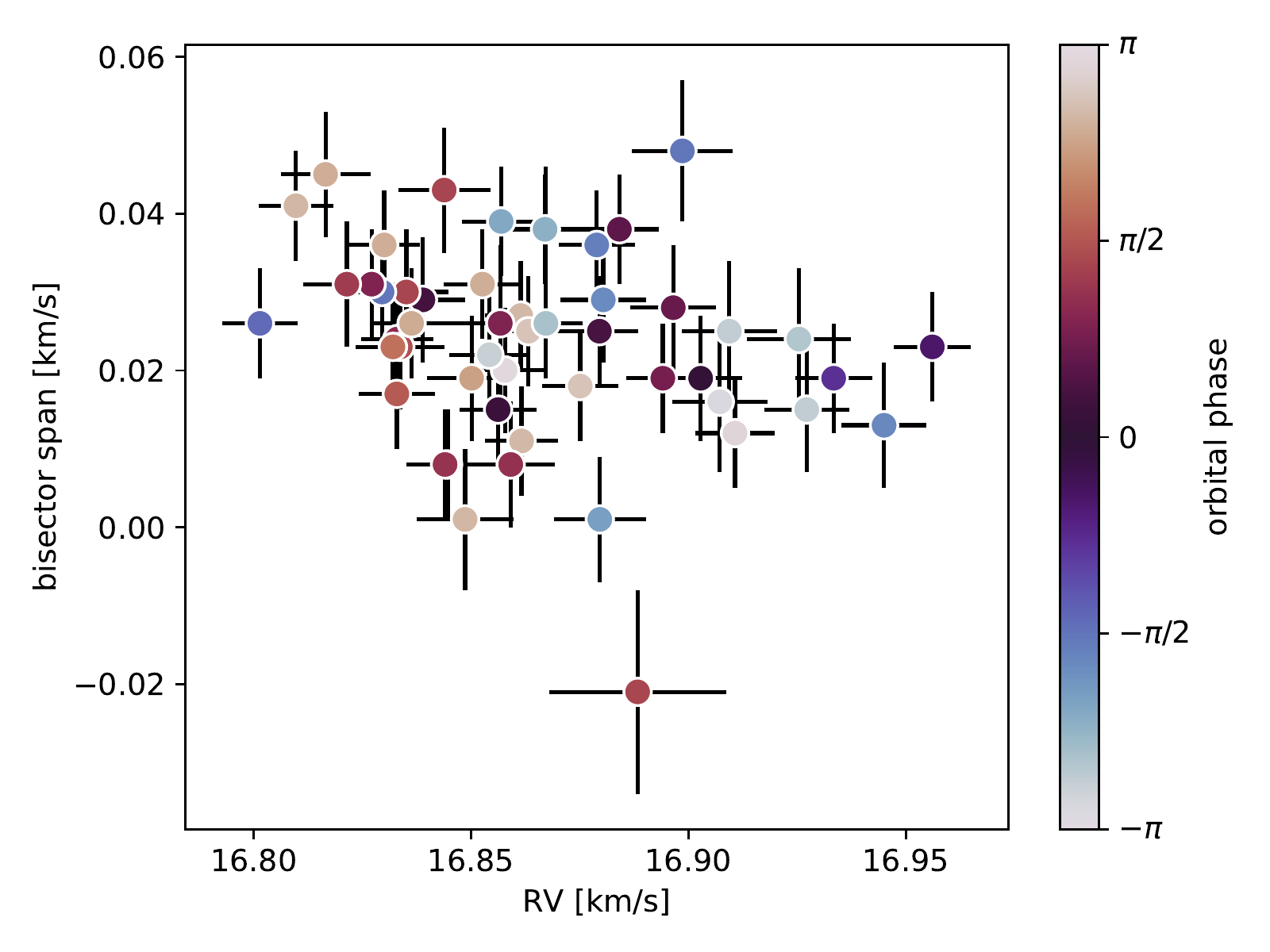}
    \includegraphics[width=1\hsize]{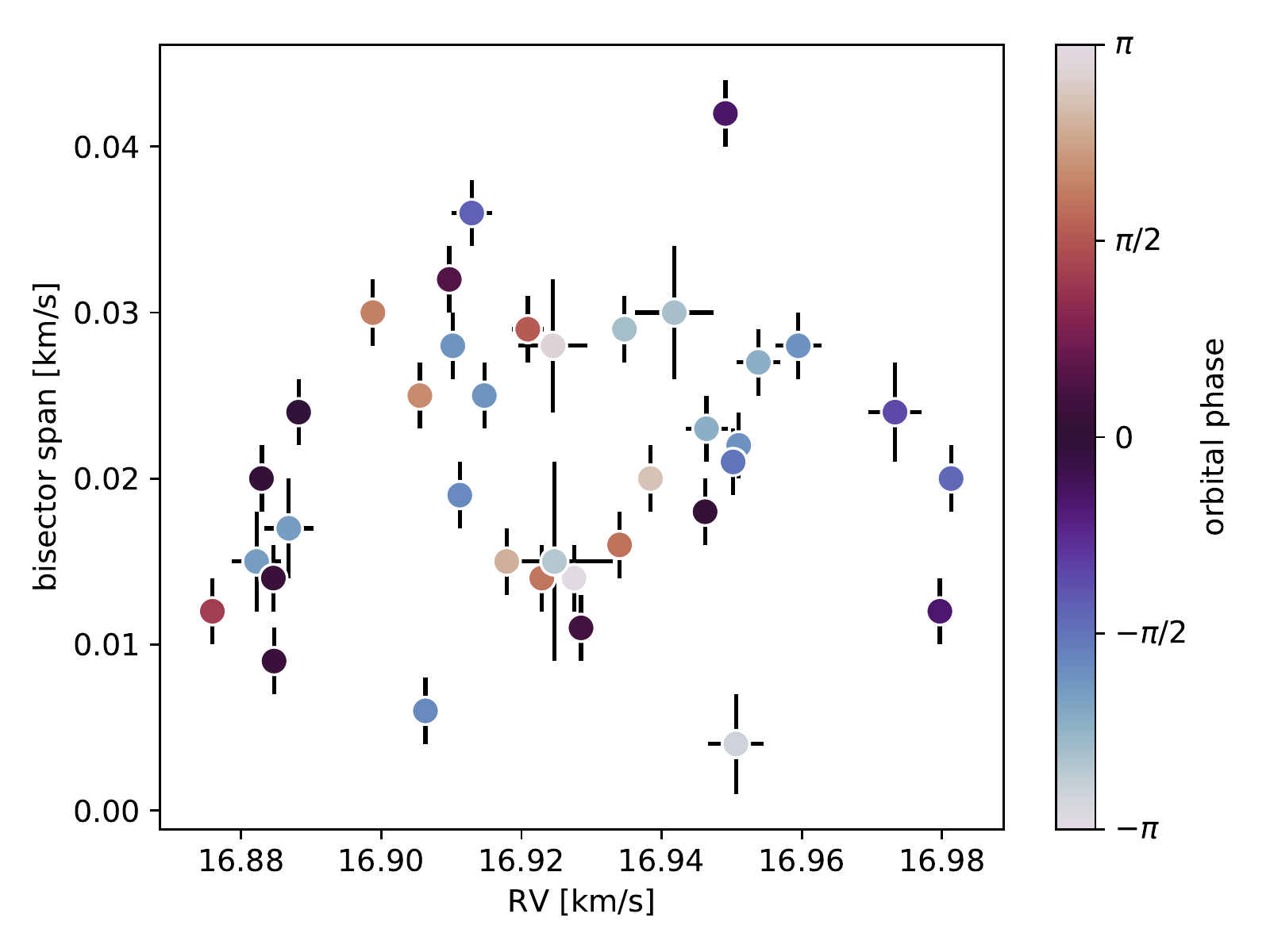}
    \caption{Bisector span as a function of radial velocity, for FEROS (top) and HARPS (bottom) data. The points are colour-coded by orbital phase. No correlations are apparent.}
    \label{fig:bis-rv-phase}
\end{figure}

Correlations between the bisector spans and the radial velocities can also indicate a stellar origin for radial velocity variations \citep[e.g.][]{Queloz01}. The radial velocity measurements are plotted against the bisector spans in Figure \ref{fig:bis-rv-phase} for both FEROS and HARPS data. Spearman correlation tests indicate no significant correlation between the bisector spans and the radial velocities, nor between the bisector spans and the orbital phases of TOI-201.01, for either data-set: For FEROS, we find a Spearman correlation coefficient $r_s=-0.31$ with $p\mathrm{\textnormal{-}value}= 0.03$ between the RVs and bisector spans, and $r_s=-0.02$ with $p\mathrm{\textnormal{-}value}= 0.09$ between the orbital phases and bisector spans; for HARPS, we find $r_s=0.15$ with $p\mathrm{\textnormal{-}value}= 0.38$ between the RVs and bisector spans, and $r_s=-0.13$ with $p\mathrm{\textnormal{-}value}= 0.44$ between the orbital phases and bisector spans.

\begin{figure}
    \centering
    \includegraphics[width=1\hsize]{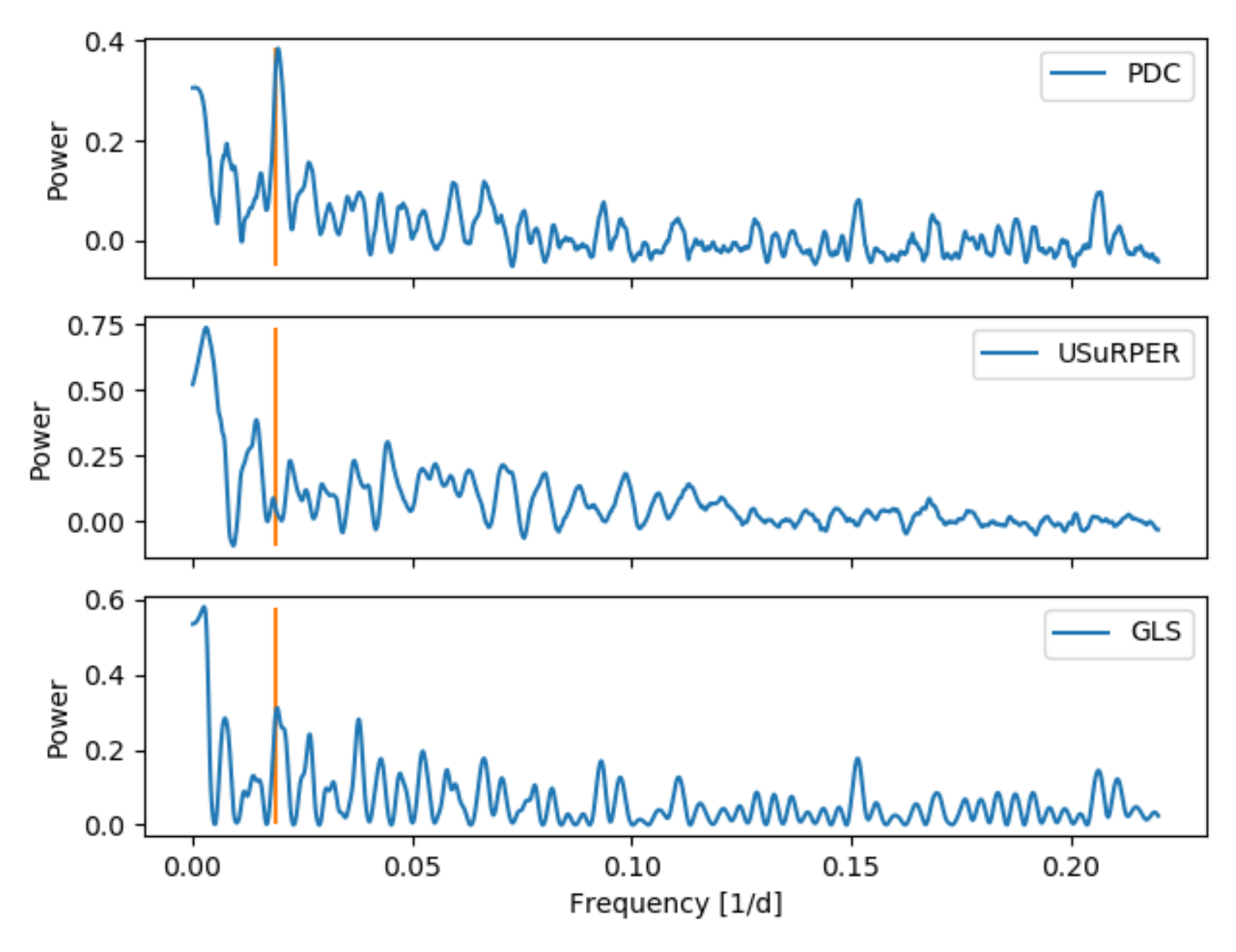}
    \caption{PDC (top), USuRPER (middle), and, for comparison, GLS (bottom) periodograms of the joint FEROS-HARPS RV time series. The vertical orange line indicates the period of the planetary candidate TOI-201.01. The peak at the planetary period is enhanced in the PDC periodogram compared to the GLS periodogram, and suppressed in the USuRPER periodogram which shows a peak in low frequencies.}
    \label{fig:periodograms-PDC}
\end{figure}

In order to further verify that the $\mathrm{\sim 53\, d}$ signal is planetary in nature, we also computed the Phase Distance Correlation (PDC) periodogram \citep{Zucker18} for the FEROS data. We find a stronger signal at the $\mathrm{53\, d}$ period in the PDC periodogram than in a comparison GLS periodogram, as expected for eccentric orbits, as shown in Fig. \ref{fig:periodograms-PDC}. We also ran the recent PDC extension, the USuRPER \citep{Binnenfeld2020}, which is designed to account for fluctuations in the entire spectral shape. A prominent peak is visible in the USuRPER low frequency region (Fig. \ref{fig:periodograms-PDC}). This result strengthens the finding that those signals are caused by some sort of periodic variations in the spectral feature shape. Both the PDC and USuRPER periodograms are implemented in the SPARTA code library \citep{SPARTA}.

We use the \texttt{juliet} python package \citep{Espinoza19juliet} to model the joint FEROS and HARPS  radial velocities. This tool allows us to jointly fit photometric data (using the \texttt{batman} package, \citealt{Kreidberg15}) and radial velocities (using the \texttt{radvel} package, \citealt{Fulton18}), and also to incorporate Gaussian Processes (via the \texttt{celerite} package, \citealt{Foreman-Mackey17}). The parameter space is explored through nested sampling, using the \texttt{MultiNest} algorithm \citep{Feroz09} in its python implementation, \texttt{PyMultiNest} \citep{Buchner14}, or the \texttt{dynesty} package \citep{Speagle20}. 

We tested six models for the radial velocities: A flat model with no Keplerian components; a single circular planet; a single eccentric planet; two eccentric planets; a single eccentric planet plus a quadratic trend to model long-term effects; and a single eccentric planet plus a Gaussian process (GP) to account for stellar activity. The full priors for each model are listed in Table \ref{tab:rvpriors}, and the posteriors in Table \ref{tab:rvpost}. For all Keplerian components, we used the periods $\mathrm{P_{01},\, P_{02}}$ and epochs $\mathrm{T0_{01},\, T0_{02}}$ listed in ExoFOP-TESS for candidates TOI-201.01 and TOI-201.02 respectively as initial constraints. For eccentric models, we use the ($\mathrm{e \sin{\omega}, \, e \cos{\omega}}$) parametrization.
 
\begin{enumerate}
    \item \label{mod:rv-no-planet} \textbf{no planet}: this model assumes that all the RV variations are due solely to jitter. The only free parameters are the systemic radial velocities and jitters for the two instruments. This model effectively provides us with a baseline log-evidence $\ln(Z) = 32.9 \pm 0.2$. 
    \item \label{mod:rv-1plc} \textbf{one-planet, circular}: in this model, we add a Keplerian with Gaussian priors for the period and T0 centred on those provided by the light curve, and an eccentricity fixed to zero. We find a surprisingly low log-evidence of $\ln(Z) = 1.16 \pm 0.02$. As can be seen in Table \ref{tab:rvpost}, the instrumental jitter values are very similar to those of the flat model, suggesting that the RV variations may be absorbed by this jitter.
    \item \label{mod:rv-1ple} \textbf{one-planet, eccentric}: in this model, we allow for free eccentricity and $\omega$ for the Keplerian. We find a much higher log-evidence of $\ln(Z) = 126.7 \pm 0.1$.
    \item \label{mod:rv-2ple} \textbf{two planets, eccentric}: In addition to the Keplerian with priors centred on the parameters of TOI-201.01, we include a second Keplerian (also with free eccentricity and $\omega$) to represent TOI-201.02. We find a decreased log-evidence with respect to model \ref{mod:rv-1ple} of $\ln(Z) = 25.96 \pm 0.01$, and the eccentricity for the 53 d Keplerian is increased.
    \item \label{mod:rv-1ple-qu} \textbf{one-planet, eccentric, plus quadratic trend}: in this model, we test the inclusion of a quadratic trend to model long-term effects. The intercept parameter is fixed to zero, as otherwise it becomes degenerate with the instrumental offsets. We find a log-evidence of $\ln(Z) = 173.60 \pm 0.01$.
    \item \label{mod:rv-1ple-gp} \textbf{one-planet, eccentric, plus GP}: in this model, we incorporate a Gaussian process, to account for stellar activity. We use an (approximate) Matern kernel, as implemented in \texttt{celerite}. We find a log-evidence of $\ln(Z) = 185.71 \pm 0.02$.
\end{enumerate}

\begin{figure*}
    \centering
    \includegraphics[width=1\hsize]{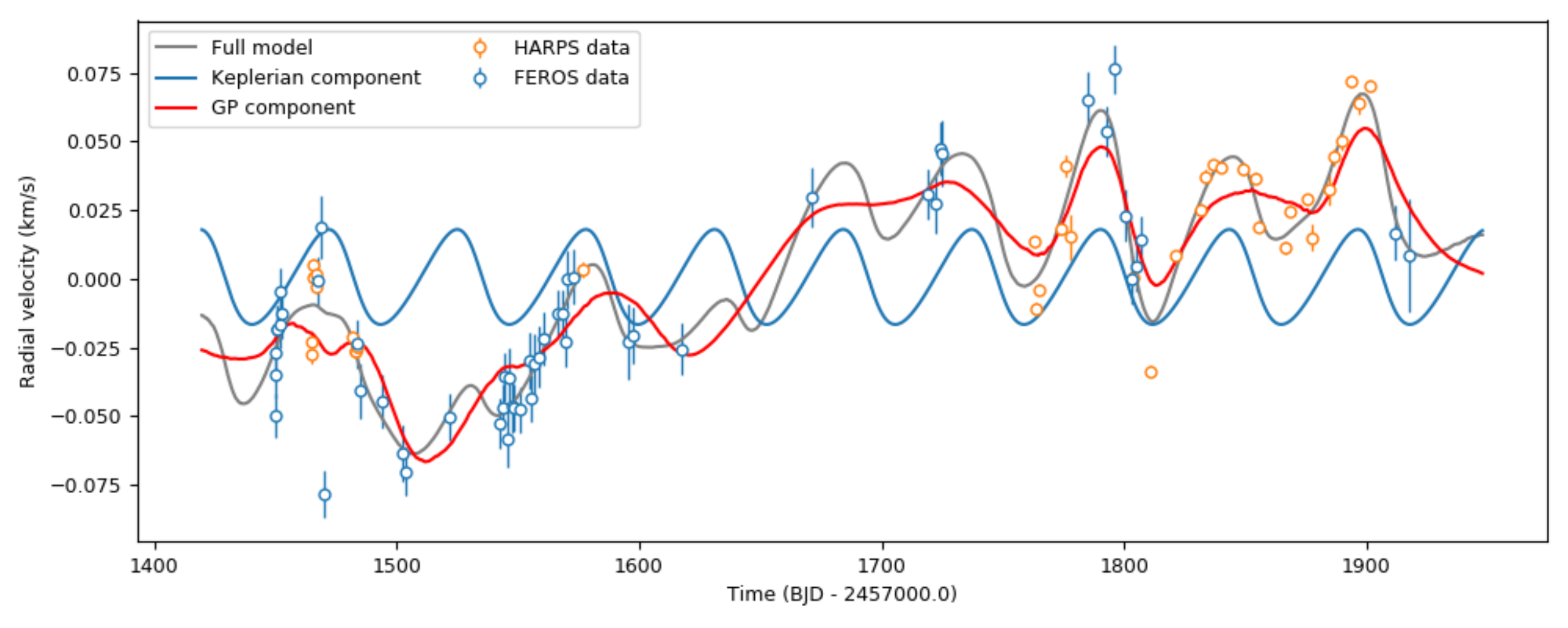}
    \caption{RV data for FEROS and HARPS, full fitted model, and Keplerian and GP model components, for the favoured model of one eccentric planet plus a Gaussian process.}
    \label{fig:rv-1pe-gp}
\end{figure*}

The model with a single eccentric Keplerian and a Gaussian process is clearly favoured, with a $\Delta \ln Z_{\ref*{mod:rv-1ple-gp}, \ref*{mod:rv-1ple}} \approx 59$ compared to the model with only a single eccentric planet, and better accounts for other trends in the data, with $\Delta \ln Z_{\ref*{mod:rv-1ple-gp}, \ref*{mod:rv-1ple-qu}} \approx 12$ compared to the model with a single eccentric planet and a quadratic trend. Likewise, the difference to the flat model is of $\Delta \ln Z_{\ref*{mod:rv-1ple-gp}, \ref*{mod:rv-no-planet}} \approx 153$. This allows us to confirm the candidate planet TOI-201.01, referred to as TOI-201 b hereafter. The full model and components, and the FEROS and HARPS RVs, are shown in Fig. \ref{fig:rv-1pe-gp}.

\begin{table}[htb]
\begin{center}
\caption{Prior parameter distributions for the RV analysis. $\mathcal{U}(a,b)$ indicates a uniform distribution between $a$ and $b$; $\mathcal{N}(a, b)$ a normal distribution with mean $a$ and standard deviation $b$; $\mathcal{J}(a,b)$ a Jeffreys or log-uniform distribution between $a$ and $b$.}
\label{tab:rvpriors}
\centering
\begin{tabular}{lll}
\hline \hline
Parameter & Distribution & Models \\
\hline
$\mu_\mathrm{{HARPS}}$ \dotfill [km/s] & $\mathcal{U}(-30,30) $ & \ref{mod:rv-no-planet}, \ref{mod:rv-1plc}, \ref{mod:rv-1ple}, \ref{mod:rv-2ple}, \ref{mod:rv-1ple-qu}, \ref{mod:rv-1ple-gp} \\
$\mu_\mathrm{{FEROS}}$ \dotfill [km/s] & $\mathcal{U}(-30,30) $ & \ref{mod:rv-no-planet}, \ref{mod:rv-1plc}, \ref{mod:rv-1ple}, \ref{mod:rv-2ple}, \ref{mod:rv-1ple-qu}, \ref{mod:rv-1ple-gp} \\
$\sigma_\mathrm{{HARPS}}$ \dotfill [km/s] & $\mathcal{J}(0.001,100) $ & \ref{mod:rv-no-planet}, \ref{mod:rv-1plc}, \ref{mod:rv-1ple}, \ref{mod:rv-2ple}, \ref{mod:rv-1ple-qu}, \ref{mod:rv-1ple-gp} \\
$\sigma_\mathrm{{FEROS}}$ \dotfill [km/s] & $\mathcal{J}(0.001,100) $ & \ref{mod:rv-no-planet}, \ref{mod:rv-1plc}, \ref{mod:rv-1ple}, \ref{mod:rv-2ple}, \ref{mod:rv-1ple-qu}, \ref{mod:rv-1ple-gp} \\
P$_\mathrm{01}$ \dotfill [d] & $\mathcal{N}(52.978306,0.1) $ & \ref{mod:rv-1plc}, \ref{mod:rv-1ple}, \ref{mod:rv-2ple}, \ref{mod:rv-1ple-qu}, \ref{mod:rv-1ple-gp} \\ 
T0$_\mathrm{01}$ \dotfill [d] & $\mathcal{N}(2458376.052124,0.1) $ & \ref{mod:rv-1plc}, \ref{mod:rv-1ple}, \ref{mod:rv-2ple}, \ref{mod:rv-1ple-qu}, \ref{mod:rv-1ple-gp} \\ 
K$_\mathrm{01}$ \dotfill [km/s] & $\mathcal{U}(0,1) $ & \ref{mod:rv-1plc}, \ref{mod:rv-1ple}, \ref{mod:rv-2ple}, \ref{mod:rv-1ple-qu}, \ref{mod:rv-1ple-gp} \\ 
$e \sin{\omega}_\mathrm{01}$ \dotfill & $\mathcal{U}(-1,1) $ & \ref{mod:rv-1ple}, \ref{mod:rv-2ple}, \ref{mod:rv-1ple-qu}, \ref{mod:rv-1ple-gp} \\
$e \cos{\omega}_\mathrm{01}$ \dotfill & $\mathcal{U}(-1,1) $ & \ref{mod:rv-1ple}, \ref{mod:rv-2ple}, \ref{mod:rv-1ple-qu}, \ref{mod:rv-1ple-gp} \\
P$_\mathrm{02}$ \dotfill [d] & $\mathcal{N}(5.849173,0.1) $ &  \ref{mod:rv-2ple} \\ 
T0$_\mathrm{02}$ \dotfill [d] & $\mathcal{N}(2458327.244194, 0.1) $ &  \ref{mod:rv-2ple} \\ 
K$_\mathrm{02}$ \dotfill [km/s] & $\mathcal{U}(0,1) $ & \ref{mod:rv-2ple} \\ 
$e \sin{\omega}_\mathrm{02}$ \dotfill & $\mathcal{U}(-1,1) $ &  \ref{mod:rv-2ple} \\
$e \cos{\omega}_\mathrm{02}$ \dotfill & $\mathcal{U}(-1,1) $ & \ref{mod:rv-2ple} \\ 
$rv_{slope}$ \dotfill [(km/s)/d] & $\mathcal{U}(-100,100) $ & \ref{mod:rv-1ple-qu} \\
$rv_{quad}$ \dotfill [(km/s)/d$^2$] & $\mathcal{U}(-100,100) $ & \ref{mod:rv-1ple-qu} \\
$\sigma_\mathrm{{GP}}$ \dotfill & $\mathcal{J}(0.01,100) $ & \ref{mod:rv-1ple-gp}\\ 
$\rho_\mathrm{{GP}}$ \dotfill & $\mathcal{J}(0.01,100) $ & \ref{mod:rv-1ple-gp}\\ 
\hline
\end{tabular}
\end{center}
\end{table}

\begin{table*}[htb]
\begin{center}
\caption{RV analysis for the different models tested. Top: posterior parameter distributions. Bottom: log-evidence and difference to the baseline no-planet model. Model \ref{mod:rv-1ple-gp} is the final adopted model.}
\label{tab:rvpost}
\centering
\resizebox{\textwidth}{!}{%
\begin{tabular}{lllllll}
\hline \hline
Parameter & Flat model (\ref{mod:rv-no-planet}) & 1 planet circ. (\ref{mod:rv-1plc}) & 1 planet ecc. (\ref{mod:rv-1ple}) & 2 planet ecc. (\ref{mod:rv-2ple}) & 1 planet + quad (\ref{mod:rv-1ple-qu}) & 1 planet + GP (\ref{mod:rv-1ple-gp}) \\
\hline
$\mu_\mathrm{{HARPS}}$ \dotfill [km/s] & $16.999^{+0.009}_{-0.038}$ & $16.92^{+0.01}_{-0.01}$ & $16.92^{+0.01}_{-0.01}$ & $17.04^{+0.04}_{-0.05}$ & $16.884^{+0.003}_{-0.004}$ & $16.91^{+0.05}_{-0.06}$ \\
$\mu_\mathrm{{FEROS}}$ \dotfill [km/s] & $16.983^{+0.010}_{-0.002}$ & $16.834^{+0.004}_{-0.003}$ & $16.875^{+0.005}_{-0.006}$ & $16.89^{+0.01}_{-0.01}$ & $16.855^{+0.003}_{-0.004}$ & $16.88^{+0.05}_{-0.06}$ \\
$\sigma_\mathrm{{HARPS}}$ \dotfill [km/s] & $0.148^{+0.007}_{-0.009}$ & $0.157^{+0.003}_{-0.003}$ & $0.068^{+0.012}_{-0.008}$ & $0.40^{+0.05}_{-0.04}$ & $0.011^{+0.002}_{-0.001}$ & $0.012^{0.001}_{0.001}$ \\
$\sigma_\mathrm{{FEROS}}$ \dotfill [km/s] & $0.189^{+0.013}_{-0.005}$ & $0.294^{+0.006}_{-0.004}$ & $0.047^{+0.005}_{-0.006}$ & $0.093^{+0.010}_{-0.008}$ & $0.020^{+0.003}_{-0.002}$ & $0.013^{0.002}_{0.002}$ \\
P$_\mathrm{01}$ \dotfill [d] & $\cdots$ & $52.787^{+0.006}_{-0.006}$ & $52.99^{+0.08}_{-0.09}$ & $53.03^{+0.04}_{-0.06}$ & $53.04^{+0.08}_{-0.09}$ & $53.04 ^{+0.05}_{-0.06}$  \\ 
T0$_\mathrm{01}$ \dotfill [d] & $\cdots$ & $2458376.130^{+0.07}_{-0.009}$ & $2458375.97^{+0.04}_{-0.05}$ & $2458376.00^{+0.05}_{-0.04}$ & $2458376.1^{+0.1}_{-0.1}$ & $2458376.11^{+0.05}_{-0.05}$  \\ 
K$_\mathrm{01}$ \dotfill [km/s] & $\cdots$ & $0.054^{+0.005}_{-0.005}$ & $0.027^{+0.014}_{-0.009}$ & $0.08^{+0.04}_{-0.04}$ & $0.026^{+0.003}_{-0.003}$ & $0.016^{+0.005}_{-0.005}$  \\ 
$e \sin{\omega}_\mathrm{01}$ \dotfill & $\cdots$ & $\cdots$ & $0.44^{+0.19}_{-0.22}$ & $0.30^{+0.25}_{-0.33}$ & $0.31^{+0.09}_{-0.09}$ & $0.20^{+0.16}_{-0.19}$ \\ 
$e \cos{\omega}_\mathrm{01}$ \dotfill & $\cdots$ & $\cdots$ & $0.21^{+0.16}_{-0.12}$ & $0.50^{+0.18}_{-0.36}$ & $0.03^{+0.09}_{-0.08}$ & $-0.10^{+0.19}_{-0.20}$ \\
$e_\mathrm{{01}}$ \dotfill & $\cdots$ & $\cdots$ & $0.52^{+0.18}_{-0.20}$ & $0.64^{+0.18}_{-0.29}$ & $0.32^{+0.09}_{-0.09}$ & $0.30^{+0.16}_{-0.13}$ \\ 
$\omega_\mathrm{{01}}$ \dotfill [deg] & $\cdots$ & $\cdots$ & $61^{+18}_{-18}$ & $33^{+26}_{-17}$ & $85^{+16}_{-14}$ & $110^{+39}_{-45}$\\
P$_\mathrm{02}$ \dotfill [d]  & $\cdots$ & $\cdots$ & $\cdots$ & $5.85^{+0.03}_{-0.05}$ & $\cdots$ & $\cdots$  \\ 
T0$_\mathrm{02}$ \dotfill [d] & $\cdots$ & $\cdots$ & $\cdots$ & $2458327.16^{+0.04}_{-0.05}$ & $\cdots$ & $\cdots$  \\ 
K$_\mathrm{02}$ \dotfill [m/s] & $\cdots$ & $\cdots$ & $\cdots$ & $0.02^{+0.02}_{-0.01}$ & $\cdots$ & $\cdots$ \\ 
$e \sin{\omega}_\mathrm{02}$ \dotfill & $\cdots$ & $\cdots$ & $\cdots$ & $-0.08^{+0.28}_{-0.19}$ & $\cdots$ & $\cdots$ \\
$\cos{\omega}_\mathrm{02}$ \dotfill & $\cdots$ & $\cdots$ & $\cdots$ & $-0.56^{+0.41}_{-0.28}$ & $\cdots$ & $\cdots$ \\ 
$e_\mathrm{02}$ \dotfill & $\cdots$ & $\cdots$ & $\cdots$ & $0.62^{+0.27}_{-0.34}$ & $\cdots$ & $\cdots$ \\ 
$\omega_\mathrm{02}$ \dotfill & $\cdots$ & $\cdots$ & $\cdots$ & $156^{+18}_{-35}$ & $\cdots$ & $\cdots$ \\
$rv_{slope}$ \dotfill [(km/s)/d] & $\cdots$ & $\cdots$ & $\cdots$ & $\cdots$ & $0.00009^{+0.00005}_{-0.00005}$ & $\cdots$ \\ 
$rv_{quad}$ \dotfill [(km/s)/d$^2$] & $\cdots$ & $\cdots$ & $\cdots$ & $\cdots$ & $0.00000013^{+0.00000011}_{-0.00000011}$ & $\cdots$ \\ 
$\sigma_\mathrm{{GP}}$ \dotfill & $\cdots$ & $\cdots$ & $\cdots$ & $\cdots$ & $\cdots$ & $0.10^{+0.04}_{-0.02}$ \\ 
$\rho_\mathrm{{GP}}$ \dotfill & $\cdots$ & $\cdots$ & $\cdots$ & $\cdots$ & $\cdots$ & $55^{+20}_{-14}$ \\ 
\hline
$\ln(Z)$ \dotfill & $32.9 \pm 0.2$ & $1.16 \pm 0.02$ & $126.7 \pm 0.1$ & $25.96 \pm 0.01$ & $173.60 \pm 0.01$ & $185.71 \pm 0.02$ \\ 
$\Delta \ln Z_{1}$ \dotfill & $\cdots$ & $\approx-32$ & $\approx 94$ & $\approx -7$ & $\approx 141$ & $\approx 153$ \\ 
\hline
\end{tabular}
}
\end{center}
\end{table*}

\begin{figure}
    \centering
    \includegraphics[width=1\hsize]{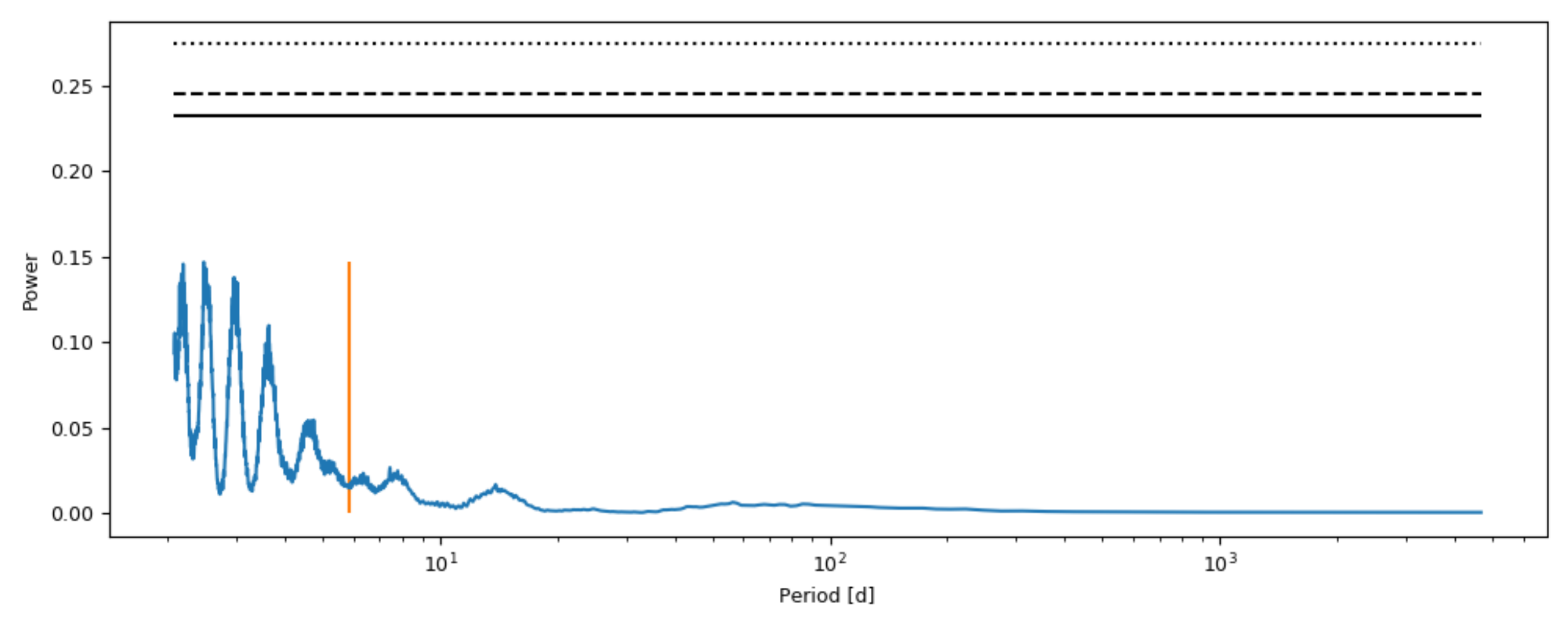}
    \includegraphics[width=1\hsize]{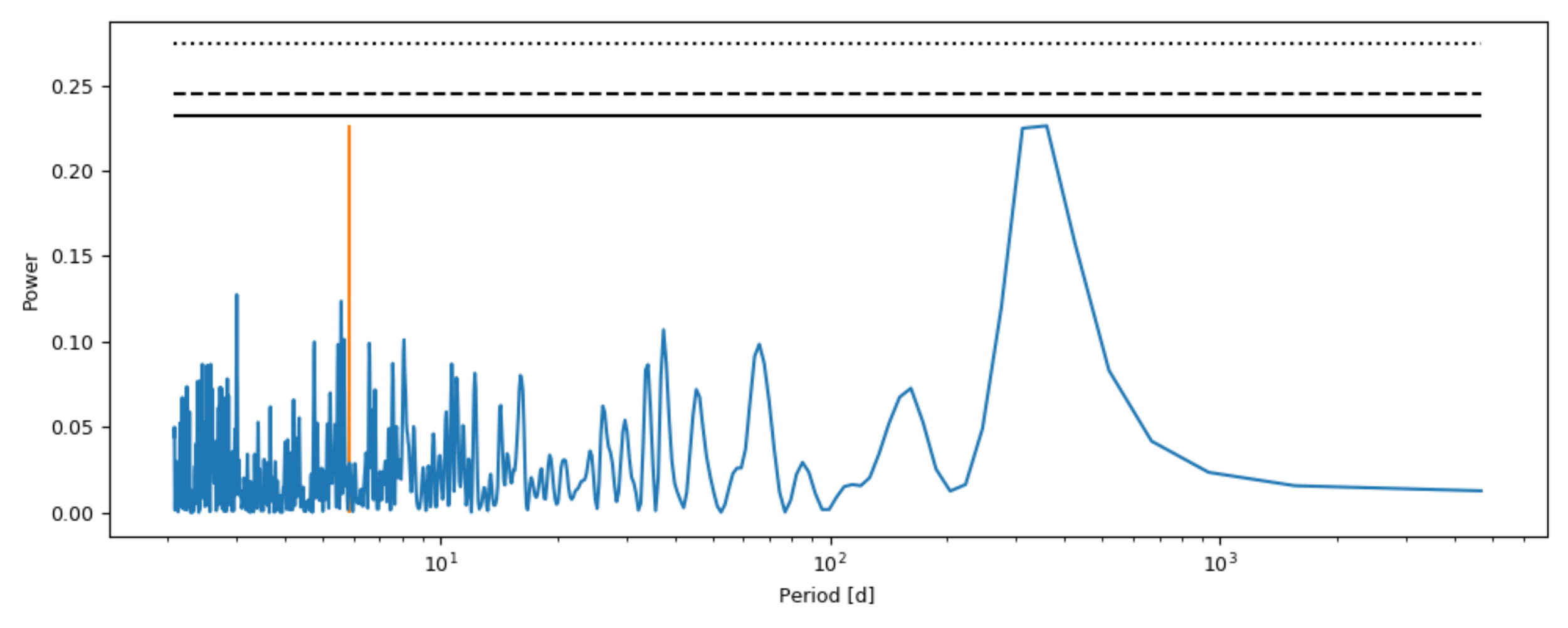}
    \includegraphics[width=1\hsize]{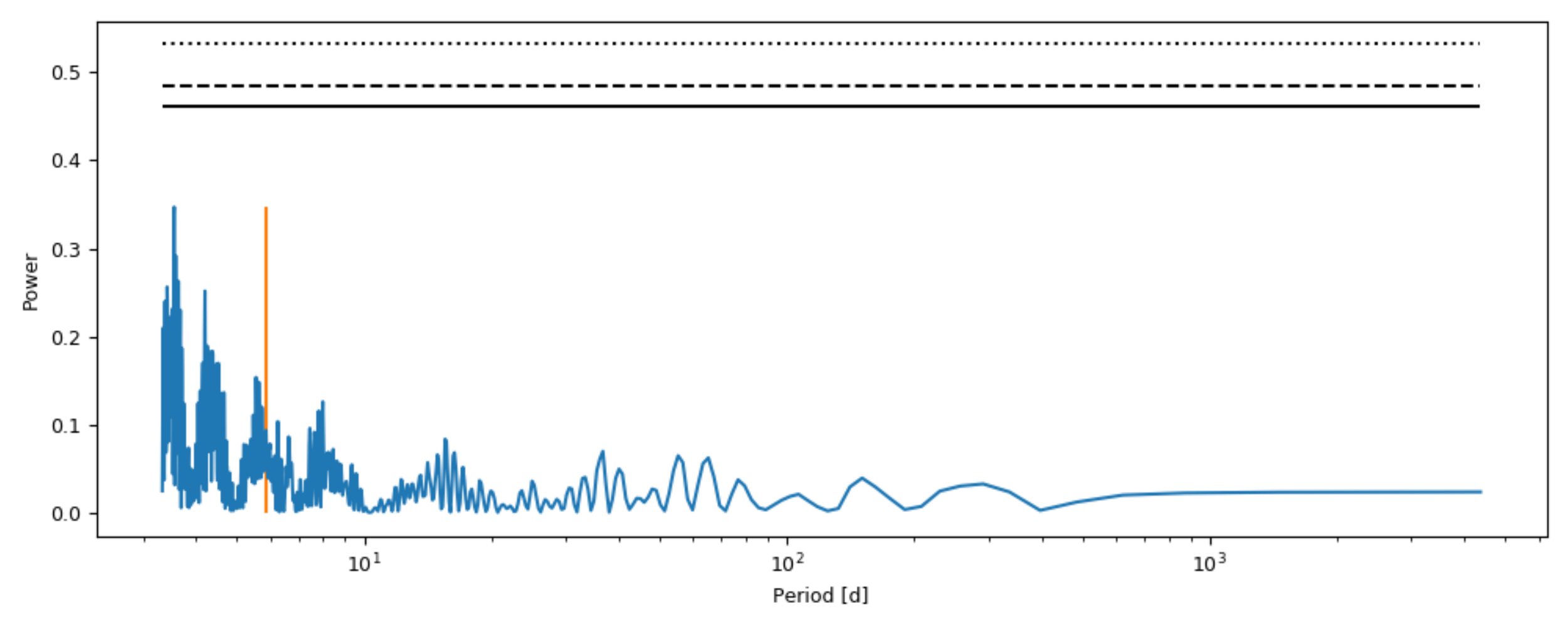}
    \caption{Periodogram of the RV residuals to models \ref{mod:rv-1ple-gp} (top) and \ref{mod:rv-1ple-qu} (middle) for the full RV dataset, and of the RV residuals to model \ref{mod:rv-1ple-gp} for the HARPS data alone (bottom). The vertical line indicates the period of the planetary candidate TOI-201.02. There are no significant signals in any of the periodograms.}
    \label{fig:rv-1pe-gp-residuals}
\end{figure}

We inspected the residuals of the RVs after subtracting model \ref{mod:rv-1ple-gp}, to check if there are any significant periodicities remaining. In particular, we wish to verify whether there is any signal at the period of the planetary candidate TOI-201.02. Figure \ref{fig:rv-1pe-gp-residuals} (top panel) shows the periodogram of the residuals, with the 5 d period of TOI-201.02 highlighted. There is no peak at or close to this period, nor any significant signals visible. We also inspected the residuals of the RVs to model \ref{mod:rv-1ple-qu}, to rule out the possibility that the signal could be absorbed by the GP. Figure \ref{fig:rv-1pe-gp-residuals} (middle panel) shows their periodogram; there are no significant signals at any period, and a peak emerges around the $\mathrm{\sim 300 \,d}$ period seen in the H$_\alpha$ and log($R^\prime_{HK}$) periodograms, suggesting this model is less efficient at suppressing the long-term activity-induced RV variations.

The question arises of whether we can expect TOI-201.02 to be detectable in RVs. We use the radius of TOI-201.02, and the revised mass-radius relations of \cite{Otegi20}, to estimate a mass of $\mathrm{M_p \approx 6.38\, M_\oplus}$ for TOI-201.02, which leads to an RV semi-amplitude of $\mathrm{K \approx 1.88\, m/s}$. While such small amplitudes are in principle detectable with HARPS, it falls below our median RV error, and the periodogram of the residuals to model \ref{mod:rv-1ple-gp} for HARPS data alone (Fig. \ref{fig:rv-1pe-gp-residuals}, bottom panel) shows no significant signals. We also tested whether such a planet would be recoverable from our data, by injecting a Keplerian into the randomly shuffled residuals. For the injected Keplerian, we assumed a circular orbit, and used the period and epoch from the transit data and the mass estimated from the radius. The periodogram of these radial velocities showed no significant signals at any period.

Although we cannot detect TOI-201.02 in the RV data, we can still use this data to estimate its maximum mass. For this, we ran a more constrained two-planet model on the FEROS and HARPS data, fixing the periods and epochs to those listed in ExoFOP-TESS. We also fixed the eccentricity of the inner planet to zero, since we can expect its orbit to be circularized by its proximity to the host star. Using the semi-amplitude of the resulting Keplerian fit, $K=1.76^{+1.84}_{-1.14}\, m/s$, we estimate a maximum mass for TOI-201.02 of $M_{max} \approx 37.5\, M_\oplus$ with 98\% confidence.

\subsection{Joint transit and radial velocity model}

Having shown that TOI-201 b can be detected in the radial velocities alone, we perform a joint modelling of the full set of radial velocities and transit data. We adopt a one-planet model, with $\mathrm{e \sin{\omega}}$ and $\mathrm{e \cos{\omega}}$ constrained using the posteriors from model \ref{mod:rv-1ple-gp}. We also include two separate GPs: one on the joint FEROS, HARPS, CORALIE, and \textsc{Minerva}-Australis radial velocities, to account for the stellar activity, as done in model \ref{mod:rv-1ple-gp}; and one on the \textit{TESS} light curve, to account for instrumental effects such as the over-estimation of flux for the partial transit in Sector 8 (Fig. \ref{fig:lightcurve}, bottom panel). Rather than fit for the planet-to-star radius ratio and impact parameter of the orbit directly, we adopt the ($\mathrm{r_1, \, r_2}$) parametrization of \cite{Espinoza18}, which allows for efficient sampling with uniform priors. 

\begin{figure*}
    \gridline{\fig{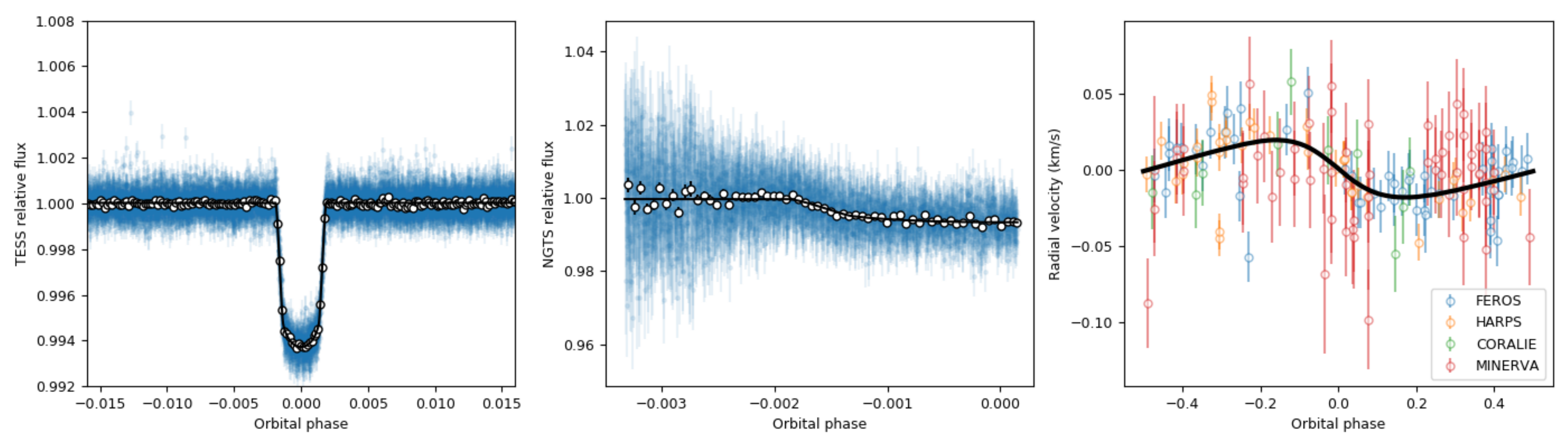}{0.9\hsize}{(a)}}
    \gridline{\fig{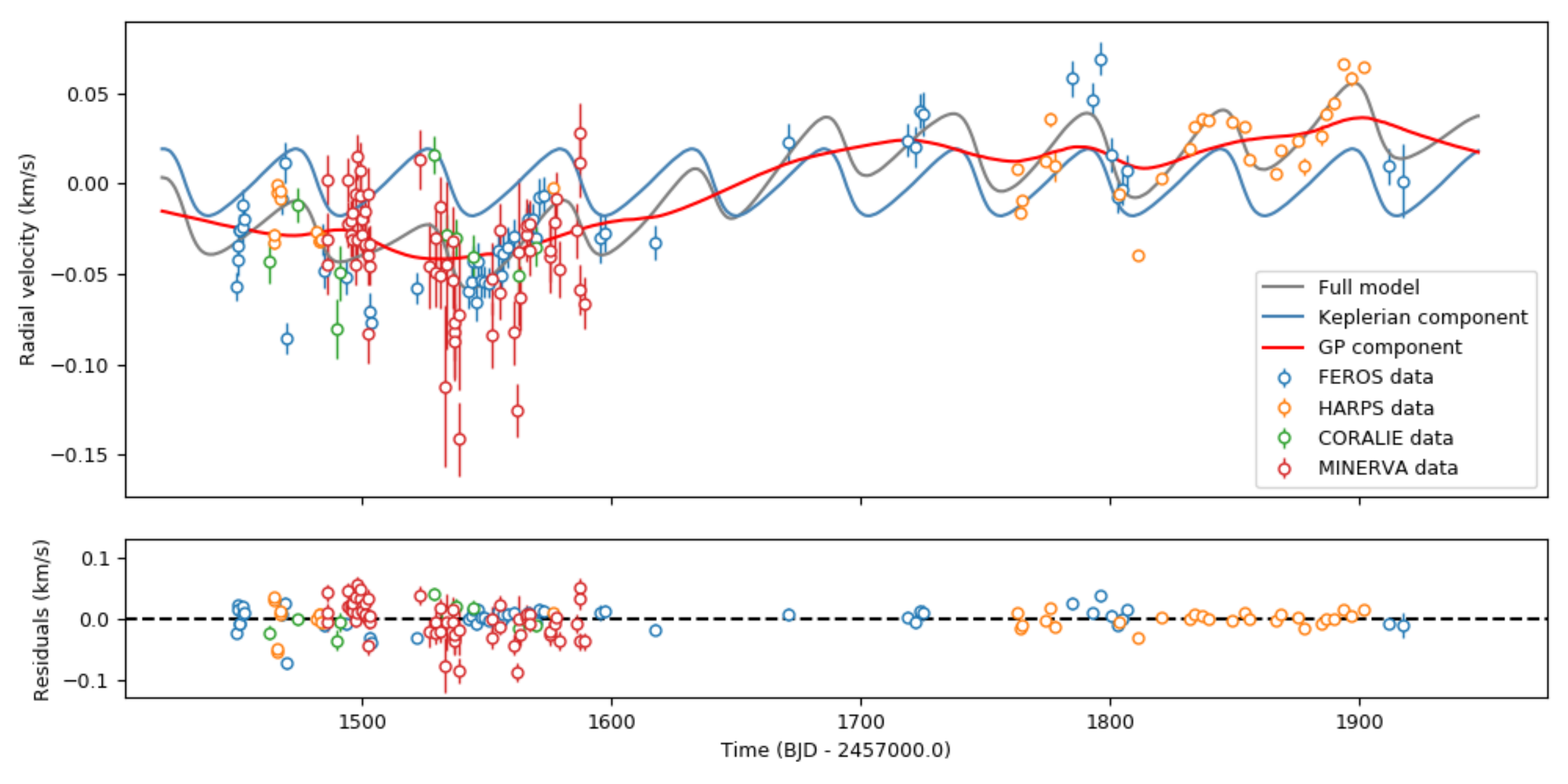}{0.9\hsize}{(b)}}
    \gridline{\fig{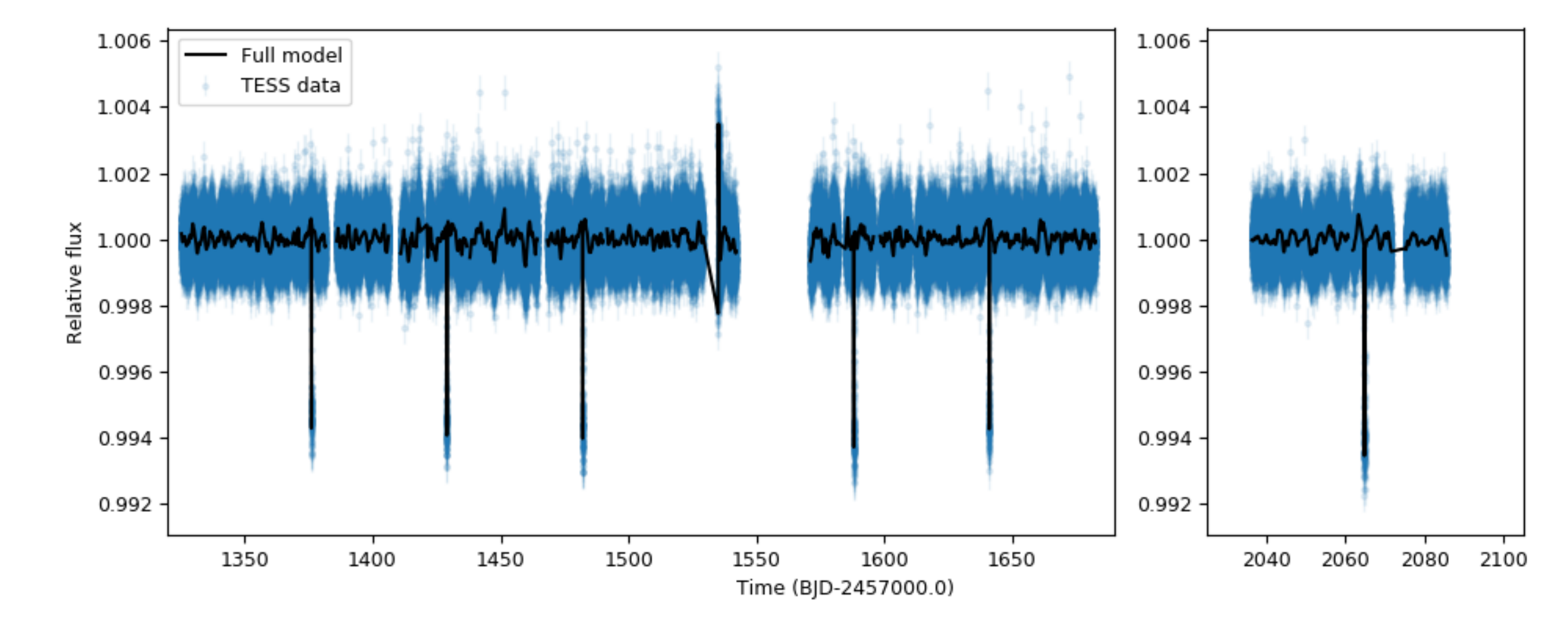}{0.9\hsize}{(c)}}
    \caption{Joint transit and RV model. (a): Left panel: phase-folded \textit{TESS} data (blue points), binned \textit{TESS} data (white points) and model (black line). Centre panel: as left, for NGTS data. Right panel: phase-folded FEROS, HARPS, CORALIE, and \textsc{Minerva}-Australis radial velocities, with the RV GP component removed, and Keplerian model component (black line). (b): Top: Radial velocity data for FEROS, HARPS, CORALIE, and \textsc{Minerva}-Australis, full model, and Keplerian and GP components. Bottom: RV residuals. (c): \textit{TESS} data and full model for Sectors 1-13 (left panel) and 27-28 (right panel).}
    \label{fig:full-model}
\end{figure*}

We find a log-evidence of $\ln(Z) = 1236042.914 \pm 0.008$. The priors are listed in Table \ref{tab:fullpriors}, and the posteriors in Table \ref{tab:fullpost}. Figure \ref{fig:full-model} shows the phase-folded data and the full model for the light curves and radial velocities. Using the stellar parameters obtained in \ref{sec:starparam}, and the posterior distributions of the full model, we compute a mass of $\mathrm{0.42^{+0.05}_{-0.03}\, M_J}$ and a radius of $\mathrm{1.008^{+0.012}_{-0.015}\, R_J}$ for TOI-201 b. The physical parameters, and derived orbital parameters, are also listed in Table \ref{tab:fullpost}.

The \textsc{Minerva}-Australis radial velocities have larger error bars and show more dispersion from the fitted model than those of the other three RV datasets, as can be clearly seen in Fig. \ref{fig:full-model} (b, bottom panel). Therefore, we also tested an analogous model in which the \textsc{Minerva}-Australis RVs were not included. 
The resulting parameters are the same within uncertainties, and the uncertainties on the parameters are similar for both models. This being the case, we prefer the model including the \textsc{Minerva}-Australis RVs.
	
\begin{table}[htb]
\begin{center}
\caption{Prior parameter distributions for the joint transit and RV analysis. $\mathcal{U}(a,b)$ indicates a uniform distribution between $a$ and $b$; $\mathcal{N}(a, b)$ a normal distribution with mean $a$ and standard deviation $b$; $\mathcal{J}(a,b)$ a Jeffreys or log-uniform distribution between $a$ and $b$.}
\label{tab:fullpriors}
\centering
\begin{tabular}{ll}
\hline \hline
Parameter & Distribution \\
\hline
$\mu_\mathrm{{HARPS}}$ \dotfill [km/s] & $\mathcal{U}(16.85,16.96)$ \\
$\mu_\mathrm{{FEROS}}$ \dotfill [km/s] & $\mathcal{U}(16.82,16.93)$  \\
$\mu_\mathrm{{CORALIE}}$ \dotfill [km/s] & $\mathcal{U}(-30,30)$  \\
$\mu_\mathrm{{MINERVA}}$ \dotfill [km/s] & $\mathcal{U}(-30,30)$  \\
$\sigma_\mathrm{{HARPS}}$ \dotfill [km/s] & $\mathcal{J}(0.011, 0.013)$  \\
$\sigma_\mathrm{{FEROS}}$ \dotfill [km/s] & $\mathcal{J}(0.011, 0.015)$  \\
$\sigma_\mathrm{{CORALIE}}$ \dotfill [km/s] & $\mathcal{J}(0.001, 100)$  \\
$\sigma_\mathrm{{MINERVA}}$ \dotfill [km/s] & $\mathcal{J}(0.001, 100)$  \\
P$_\mathrm{b}$ \dotfill [d] & $\mathcal{N}(52.978306,0.1)$  \\ 
T0$_\mathrm{b}$ \dotfill [d] & $\mathcal{N}(2458376.052124,0.1)$  \\ 
K$_\mathrm{b}$ \dotfill [km/s] & $\mathcal{U}(0.011,0.021)$  \\ 
$e \sin{\omega}_\mathrm{b}$ \dotfill & $\mathcal{U}(0.01,0.36)$  \\
$e \cos{\omega}_\mathrm{b}$ \dotfill & $\mathcal{U}(-0.30,0.09)$  \\
$\sigma_\mathrm{{GP, RV}}$ \dotfill & $\mathcal{J}(0.01,100)$ \\ 
$\rho_\mathrm{{GP, RV}}$ \dotfill & $\mathcal{J}(0.01,100)$ \\ 
r$_\mathrm{{1,b}}$ \dotfill & $\mathcal{U}(0,1)$ \\ 
r$\mathrm{_{2,b}}$ \dotfill & $\mathcal{U}(0,1)$ \\ 
$\rho$ \dotfill & $\mathcal{J}(100., 10000.)$ \\ 
q$_\mathrm{{1,TESS}}$ \dotfill & $\mathcal{U}(0,1)$ \\ 
q$_\mathrm{{2,TESS}}$ \dotfill & $\mathcal{U}(0,1)$ \\ 
$m_\mathrm{{d,TESS}}$ \dotfill & 1.0 (fixed) \\
$m_\mathrm{{flux,TESS}}$ \dotfill & $\mathcal{N}(0,0.1)$ \\ 
$\sigma_\mathrm{{TESS}}$ \dotfill & $\mathcal{J}(0.1,1000)$ \\ 
$\sigma_\mathrm{{GP, TESS}}$ \dotfill & $\mathcal{J}(1\times10^{-6},1000000)$ \\ 
$\rho_\mathrm{{GP, TESS}}$ \dotfill & $\mathcal{J}(0.001,1000)$ \\ 
q$_\mathrm{{1,NGTS}}$ \dotfill & $\mathcal{U}(0,1)$ \\ 
q$_\mathrm{{2,NGTS}}$ \dotfill & $\mathcal{U}(0,1)$ \\ 
$m_\mathrm{{d,NGTS}}$ \dotfill & 1.0 (fixed) \\
$m_\mathrm{{flux,NGTS}}$ \dotfill & $\mathcal{N}(0,0.1)$ \\ 
$\sigma_\mathrm{{NGTS}}$  \dotfill & $\mathcal{J}(0.1,1000)$ \\ 
\hline
\end{tabular}
\end{center}
\end{table}

\begin{table}[htb]
\begin{center}
\caption{Parameters for planet TOI-201 b. \textit{Top}: Posterior parameter distributions for the joint transit and RV analysis. \textit{Bottom}: derived orbital parameters and physical parameters.}
\label{tab:fullpost}
\centering
\begin{tabular}{ll}
\hline \hline
Parameter & Distribution \\
\hline
$\mu_\mathrm{{HARPS}}$ \dotfill [km/s] & $16.92^{+0.01}_{-0.01}$ \\
$\mu_\mathrm{{FEROS}}$ \dotfill [km/s] & $16.89^{+0.01}_{-0.01}$  \\
$\mu_\mathrm{{CORALIE}}$ \dotfill [km/s] & $16.87^{+0.02}_{-0.01}$  \\
$\mu_\mathrm{{MINERVA}}$ \dotfill [km/s] & $0.02^{+0.01}_{-0.01}$  \\
$\sigma_\mathrm{{HARPS}}$ \dotfill [km/s] &  $0.0117^{+0.0008}_{-0.0005}$ \\
$\sigma_\mathrm{{FEROS}}$ \dotfill [km/s] & $0.014^{+0.001}_{-0.001}$  \\
$\sigma_\mathrm{{CORALIE}}$ \dotfill [km/s] & $0.019^{+0.009}_{-0.007}$  \\
$\sigma_\mathrm{{MINERVA}}$ \dotfill [km/s] & $0.026^{+0.004}_{-0.004}$  \\
P$_\mathrm{b}$ \dotfill [d] & $52.97818^{+0.00004}_{-0.00004}$  \\ 
T0$_\mathrm{b}$ \dotfill [BJD] & $2458376.0520^{+0.0003}_{-0.0003}$  \\ 
K$_\mathrm{b}$ \dotfill [km/s] & $0.019^{+0.001}_{-0.002}$  \\ 
$e \sin{\omega}_\mathrm{b}$ \dotfill & $0.28^{+0.06}_{-0.09}$ \\
$e \cos{\omega}_\mathrm{b}$ \dotfill & $0.04^{+0.04}_{-0.07}$ \\
$\sigma_\mathrm{{GP}}$ \dotfill & $0.025^{+0.008}_{-0.006}$ \\ 
$\rho_\mathrm{{GP}}$ \dotfill & $70^{+21}_{-27}$ \\ 
r$_\mathrm{{1,b}}$ \dotfill & $0.82^{+0.01}_{-0.01}$ \\ 
r$_\mathrm{{2,b}}$ \dotfill & $0.0786^{+0.0010}_{-0.0007}$ \\ 
$\rho$ \dotfill & $854^{+288}_{-154}$ \\ 
q$_\mathrm{{1,TESS}}$ \dotfill & $0.22^{+0.06}_{-0.06}$ \\ 
q$_\mathrm{{2,TESS}}$ \dotfill & $0.25^{+0.33}_{-0.18}$ \\ 
$m_\mathrm{{d,TESS}}$ \dotfill & 1.0 \\
$m_\mathrm{{flux,TESS}}$ \dotfill & $-0.00002^{+0.00002}_{-0.00002}$ \\ 
$\sigma_\mathrm{{TESS}}$ \dotfill & $73^{+5}_{-5}$ \\ 
$\sigma_\mathrm{{GP, TESS}}$ \dotfill & $0.00030^{+0.00001}_{-0.00001}$ \\ 
$\rho_\mathrm{{GP, TESS}}$ \dotfill & $0.64^{+0.03}_{-0.03}$ \\ 
q$_\mathrm{{1,NGTS}}$ \dotfill & $0.47^{+0.23}_{-0.19}$ \\ 
q$_\mathrm{{2,NGTS}}$ \dotfill & $0.38^{+0.32}_{-0.25}$ \\ 
$m_\mathrm{{d,NGTS}}$ \dotfill & 1.0 \\
$m_\mathrm{{flux,NGTS}}$ \dotfill & $0.0001^{+0.0002}_{-0.0002}$ \\ 
$\sigma_\mathrm{{NGTS}}$ \dotfill & $5^{+56}_{-4}$ \\ 
\hline
$e_\mathrm{b}$ \dotfill &  $0.28^{+0.06}_{-0.09}$ \\
$\omega_\mathrm{b}$ \dotfill [deg] & $82^{+14}_{-9}$ \\
$a$ \dotfill [AU] & $0.30^{+0.02}_{-0.03}$ \\
$\mathrm{M_b}$ \dotfill [M$_\mathrm{J}$] & $0.42^{+0.05}_{-0.03}$ \\
$\mathrm{R_b}$ \dotfill [R$_\mathrm{J}$] & $1.008^{+0.012}_{-0.015}$ \\
$T_\mathrm{eq}$\footnote{Time-averaged equilibrium temperature, using Eq. 16 of \cite{Mendez17} assuming bond albedo $A=0$, heat distribution $\beta=0.5$, and emissivity $\epsilon=1$.} \dotfill [K] & $759^{+37}_{-26}$ \\
\hline
\end{tabular}
\end{center}
\end{table}

\section{Discussion and Conclusions} \label{sec:disc}

We have presented the confirmation through radial velocities of the transiting planet TOI-201 b. In addition to the RV signal due to TOI-201 b, we also see long-period signals in both the radial velocities and in the H$_\alpha$ and log($R^\prime_{HK}$) activity indicators. We therefore conclude that the long-period variations in RV are likely due to stellar activity, and model them using a Gaussian process. 

We are unable to confirm the second planetary candidate around this star, TOI-201.02. Two-planet fits to the radial velocities are not favoured by likelihood comparison, and there are no signals at $\sim 5.8$ d in the RV residuals to the one-planet fit. Likewise, we were not able to retrieve a Keplerian, modelled using the candidate's period and estimated mass, that we injected into the randomly shuffled residuals of the one-planet model.

\begin{figure*}[htb]
    \centering
    \includegraphics[width=1\hsize]{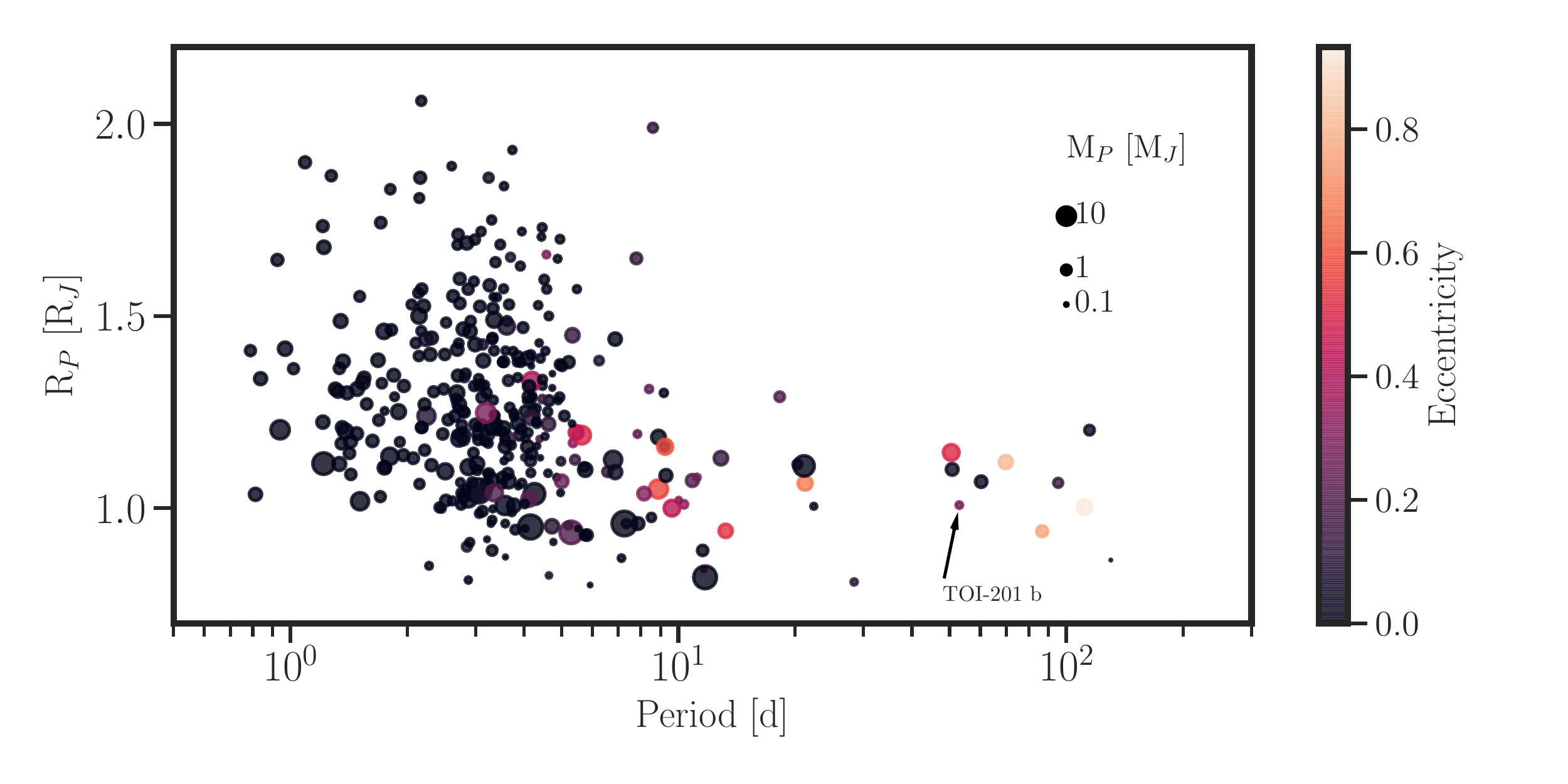}
    \caption{Radius-period diagram for giant planets ($\mathrm{R_p \geq 0.8 R_J}$), with masses and radii measured to better than 25\%, obtained from the TEPCAT catalogue \citep{Southworth11}. The markers are scaled by planet mass and colour-coded by eccentricity. TOI-201 b is labelled.}
    \label{fig:giant-planets-radius-period}
\end{figure*}

TOI-201 b is an eccentric warm giant, orbiting a young F-type star. More specifically, TOI-201 falls within the youngest 5\% of exoplanet host stars with measured ages\footnote{According to the ages reported in the NASA Exoplanet Archive, located at \url{https://exoplanetarchive.ipac.caltech.edu/index.html}}, making this system a valuable addition to the known planets around young stars, which are important for testing and constraining planet formation and evolution theories \citep[e.g.][]{Baraffe08, Mordasini12, Mordasini12b}. At $\lesssim 1 \,\mathrm{Gyr}$, some evolutionary processes that are typically not observable are still ongoing, such as photoevaporation of planetary envelopes \citep{David20} for small planets. Likewise, giant planet formation theory has difficulties fixing the luminosity of a cooling planet as a function of time ("hot start" vs. "cool start" models, e.g. \citealt{Mordasini12}), and representatives of giant planets that are currently cooling may help to constrain that. TOI-201 b also joins the small but growing population of longer-period giant planets, helping to populate a still relatively sparse region of the radius-period diagram (Figure \ref{fig:giant-planets-radius-period}). On this diagram, it is located closest to Kepler-117 c (period $\mathrm{50.790391 \pm 0.000014 \, d}$, radius $\mathrm{1.101 \pm 0.035 \, R_J}$, mass $\mathrm{1.84 \pm 0.18 \, M_J}$, eccentricity $\mathrm{0.0323 \pm 0.0033}$, from \citealt{Kepler117}) and Kepler-30 c (period $\mathrm{60.32503 \pm 0.00010 \, d}$, radius $\mathrm{1.069 \pm 0.025 \, R_J}$, mass $\mathrm{1.686 \pm 0.016 \, M_J}$, eccentricity $\mathrm{0.0115 \pm 0.0005}$, from \citealt{Kepler30}) but is noticeably less massive and dense, and more eccentric, than both these planets. Given the relative youth of the TOI-201 system, with a stellar age of $\mathrm{0.87^{+0.46}_{-0.49} \, Gyr}$ (for comparison, \citealt{Kepler117} find an age of $\mathrm{5.3 \pm 1.4 \, Gyr}$ for Kepler-117), the difference in density may be partly explained by a still-contracting TOI-201 b. 

\begin{figure}[htb]
    \centering
    \includegraphics[width=1\hsize]{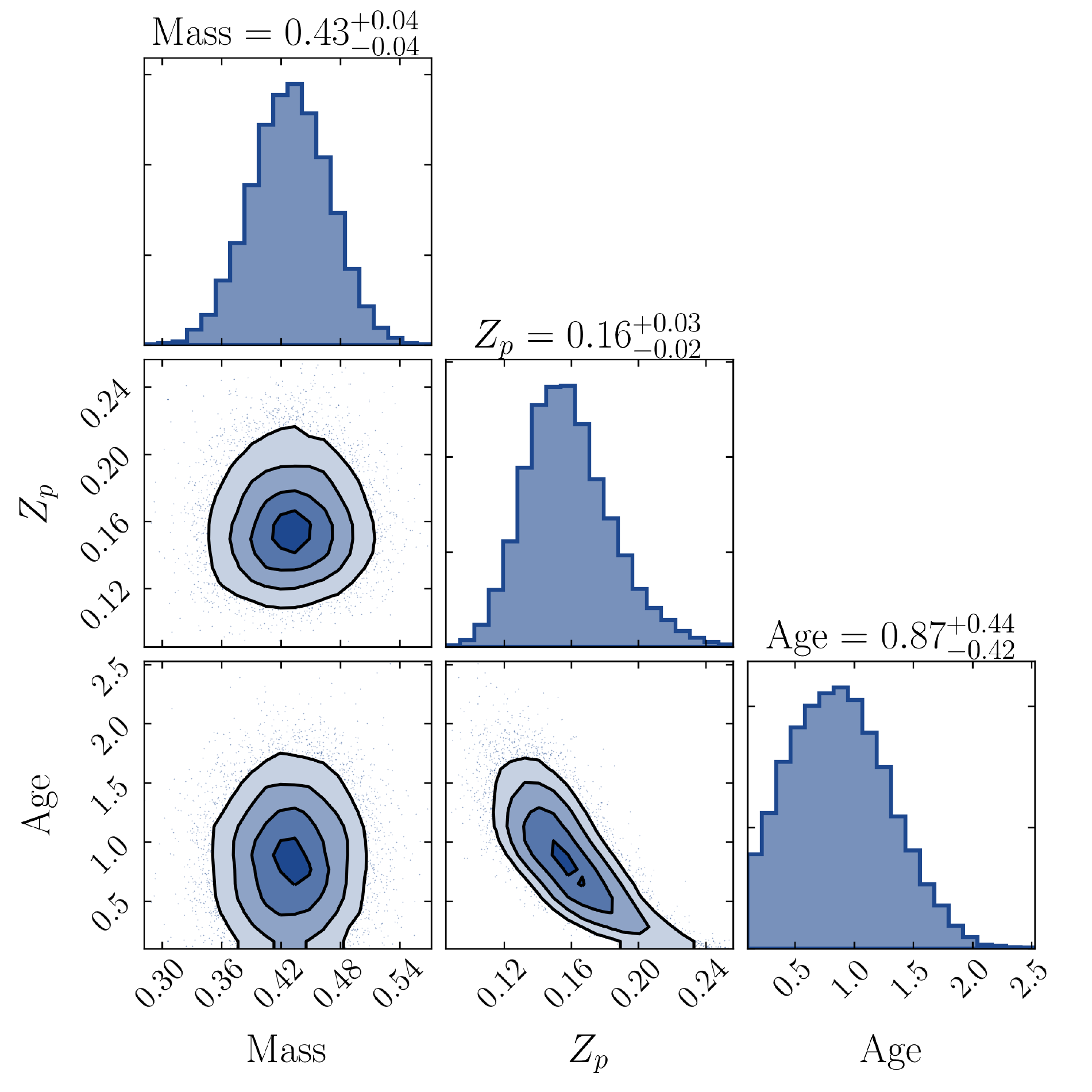}
    \caption{Posterior distribution of heavy elements, age, and mass, from the structure models of \cite{Thorngren19}.}
    \label{fig:heavy-elements}
\end{figure}

As noted in the introduction, the size and mass of warm Jupiters is determined by their metallicity. We use the structure models of \cite{Thorngren19} to constrain the heavy element mass for TOI-201 b. 
Using the planet mass and insolation flux, and the stellar age, we employ a retrieval algorithm to recover the metallicity which would match the observed radius.
The metallicity required to explain the planet's radius is strongly correlated with the age, indicating that it appears to still be cooling - and therefore contracting - fairly rapidly. The bulk metallicity we obtain is $Z_p = 0.16^{+0.03}_{-0.02}$, which is moderately low for the planet's mass compared to similarly massive planets from \cite{Thorngren19}, but still within the range of values obtained in that work. Given the high precision on the planetary parameters, modelling uncertainty (e.g. in the equation of state) is probably larger than the statistical uncertainty from the parameters. We estimate this at around a 20\% relative uncertainty, though recent work by \cite{Muller20} cautions that the interplay between various model assumptions (particularly in the choice of equations of state, the distribution of heavy elements, and the modelling of its effect on the opacity) may mean it is higher. The posterior distribution of the parameters is shown in Fig. \ref{fig:heavy-elements}. 

The Rossiter-McLaughlin effect \citep{Rossiter24, McLaughlin24} allows the stellar obliquity to be measured. These measurements are useful for constraining migration scenarios, and are particularly valuable for planets such as TOI-201 b, which are located at longer orbital distances where tidal realignment is no longer in play (see e.g. \citealt{Triaud18} for a review). It is worth noting that all the RV measurements used in the analysis in Section \ref{sec:analysis} fall outside the transit, and so are not affected by the Rossiter-McLaughlin effect. For TOI-201 b, the expected semi-amplitude of the Rossiter-McLaughlin signal is of $\mathrm{\approx 40 m/s}$ for an aligned orbit (using the equation of \citealt{Winn10}), making this system a good candidate for such observations.

\acknowledgments
This paper includes data collected by the \textit{TESS} mission, which are publicly available from the Mikulski Archive for Space Telescopes (MAST). Funding for the \textit{TESS} mission is provided by NASA's Science Mission directorate. 

This research has made use of the Exoplanet Follow-up Observation Program website, which is operated by the California Institute of Technology, under contract with the National Aeronautics and Space Administration under the Exoplanet Exploration Program.
We acknowledge the use of public TESS Alert data from the pipelines at the TESS Science Office and at the TESS Science Processing Operations Center.

Resources supporting this work were provided by the NASA High-End Computing (HEC) Program through the NASA Advanced Supercomputing (NAS) Division at Ames Research Center for the production of the SPOC data products. 

MH, AJ and RB acknowledge support from ANID – Millennium Science Initiative – ICN12\_009. AJ acknowledges additional support from FONDECYT project 1171208.
T.H. acknowledges support from the European Research Council under the
Horizon 2020 Framework Program via the ERC Advanced Grant Origins 83 24 28.
TD acknowledges support from MIT's Kavli Institute as a Kavli postdoctoral fellow.
SZ acknowledges support by the ISRAEL SCIENCE FOUNDATION (grant No. 848/16) and also partial support by the Ministry of Science, Technology and Space, Israel.

We thank the Swiss National Science Foundation (SNSF) and the Geneva University for their continuous support to our planet search programmes. This work has been in particular carried out in the frame of the National Centre for Competence in Research ‘PlanetS’ supported by the Swiss National Science Foundation (SNSF).

{\textsc{Minerva}}-Australis is supported by Australian Research Council LIEF Grant LE160100001, Discovery Grant DP180100972, Mount Cuba Astronomical Foundation, and institutional partners University of Southern Queensland, UNSW Sydney, MIT, Nanjing University, George Mason University, University of Louisville, University of California Riverside, University of Florida, and The University of Texas at Austin.
We respectfully acknowledge the traditional custodians of all lands throughout Australia, and recognise their continued cultural and spiritual connection to the land, waterways, cosmos, and community. We pay our deepest respects to all Elders, ancestors and descendants of the Giabal, Jarowair, and Kambuwal nations, upon whose lands the {\textsc{Minerva}}-Australis facility at Mt Kent is situated.

%

\facilities{\textit{TESS}, FEROS/MPG2.2m, HARPS/ESO3.6m, CORALIE, \textsc{Minerva}-Australis, NGTS, HRCam/SOAR}


\software{CERES \citep{Brahm17CERES}, 
          juliet \citep{Espinoza19juliet},
          ZASPE~\citep{Brahm17ZASPE},
          radvel~\citep{Fulton18},
          emcee~\citep{Foreman-Mackey13},
          MultiNest~\citep{Feroz09},
          PyMultiNest~\citep{Buchner14},
          batman~\citep{Kreidberg15},
          celerite~\citep{Foreman-Mackey17}
          }


\newpage

\appendix

\section{Radial velocity and activity indices}
\restartappendixnumbering

In this appendix, we present the radial velocities and activity indices, where applicable. Table \ref{tab:feros-data} shows the data for FEROS, and Table \ref{tab:harps-data} the data for HARPS, both computed using the \texttt{CERES} pipeline. Table \ref{tab:coralie-data} shows the radial velocities obtained from the CORALIE spectra. Table \ref{tab:minerva-data} shows the radial velocities obtained from the \textsc{Minerva}-Australis spectra. 

\begin{table*}[pht]
\begin{center}
\caption{RV and activity indices obtained from the FEROS spectra.}
\label{tab:feros-data}
\centering
\resizebox{\textwidth}{!}{%
\begin{tabular}{llllllllll}
\hline  \hline
BJD - 2457000 [d] & RV [km/s] & Bisector & FWHM & SNR & H$_\alpha$ & log($R^\prime_{HK}$) & Na~II & He~I \\
\hline
$2458449.72902056$ &   $16.8300 \pm 0.0082$ & $0.036 \pm 0.007$ & $15.5668$ & $193$ & $0.1075 \pm 0.0009$ & $-4.749 \pm 0.026$ & $0.3978 \pm 0.0020$ & $0.4992 \pm 0.0025$ \\
$2458449.82501961$ &   $16.8526 \pm 0.0090$ & $0.031 \pm 0.007$ & $15.6119$ & $168$ & $0.1076 \pm 0.0010$ & $-4.726 \pm 0.026$ & $0.3912 \pm 0.0022$ & $0.5022 \pm 0.0028$ \\
$2458449.81907519$ &   $16.8447 \pm 0.0086$ & $0.008 \pm 0.007$ & $15.6457$ & $182$ & $0.1065 \pm 0.0009$ & $-4.741 \pm 0.027$ & $0.3910 \pm 0.0021$ & $0.5019 \pm 0.0026$ \\
$2458450.78211913$ &   $16.8616 \pm 0.0084$ & $0.011 \pm 0.007$ & $15.6056$ & $187$ & $0.1070 \pm 0.0009$ & $-4.757 \pm 0.027$ & $0.3982 \pm 0.0020$ & $0.5033 \pm 0.0026$ \\
$2458450.77751112$ &   $16.8614 \pm 0.0083$ & $0.027 \pm 0.007$ & $15.6184$ & $191$ & $0.1064 \pm 0.0009$ & $-4.739 \pm 0.026$ & $0.3979 \pm 0.0020$ & $0.5012 \pm 0.0026$ \\
$2458451.80088761$ &   $16.8751 \pm 0.0087$ & $0.018 \pm 0.007$ & $15.6244$ & $179$ & $0.1087 \pm 0.0010$ & $-4.705 \pm 0.025$ & $0.3923 \pm 0.0021$ & $0.5011 \pm 0.0027$ \\
$2458451.80550488$ &   $16.8632 \pm 0.0087$ & $0.025 \pm 0.007$ & $15.6687$ & $177$ & $0.1090 \pm 0.0010$ & $-4.756 \pm 0.027$ & $0.3908 \pm 0.0021$ & $0.4983 \pm 0.0027$ \\
$2458452.81008207$ &   $16.8672 \pm 0.0093$ & $0.038 \pm 0.008$ & $15.6380$ & $161$ & $0.1093 \pm 0.0011$ & $-4.714 \pm 0.026$ & $0.4017 \pm 0.0024$ & $0.4986 \pm 0.0030$ \\
$2458467.81642493$ &   $16.8789 \pm 0.0088$ & $0.036 \pm 0.007$ & $15.5752$ & $174$ & $0.1081 \pm 0.0010$ & $-4.715 \pm 0.026$ & $0.3830 \pm 0.0022$ & $0.4967 \pm 0.0026$ \\
$2458468.80955367$ &   $16.8986 \pm 0.0116$ & $0.048 \pm 0.009$ & $15.6108$ & $115$ & $0.1101 \pm 0.0016$ & $-4.771 \pm 0.033$ & $0.3838 \pm 0.0035$ & $0.5044 \pm 0.0041$ \\
$2458469.81379926$ &   $16.8014 \pm 0.0086$ & $0.026 \pm 0.007$ & $15.7449$ & $184$ & $0.1051 \pm 0.0010$ & $-4.764 \pm 0.028$ & $0.3908 \pm 0.0021$ & $0.5026 \pm 0.0026$ \\
$2458483.80487685$ &   $16.8562 \pm 0.0088$ & $0.015 \pm 0.007$ & $15.6195$ & $174$ & $0.1103 \pm 0.0010$ & $-4.734 \pm 0.027$ & $0.3971 \pm 0.0022$ & $0.5024 \pm 0.0028$ \\
$2458484.68236998$ &   $16.8389 \pm 0.0098$ & $0.029 \pm 0.008$ & $15.6311$ & $147$ & $0.1139 \pm 0.0012$ & $-4.728 \pm 0.028$ & $0.3916 \pm 0.0025$ & $0.4989 \pm 0.0032$ \\
$2458493.74536456$ &   $16.8351 \pm 0.0096$ & $0.030 \pm 0.008$ & $15.6102$ & $152$ & $0.1136 \pm 0.0012$ & $-4.770 \pm 0.030$ & $0.3987 \pm 0.0025$ & $0.4949 \pm 0.0032$ \\
$2458502.71190832$ &   $16.8165 \pm 0.0103$ & $0.045 \pm 0.008$ & $15.6693$ & $136$ & $0.1140 \pm 0.0014$ & $-4.760 \pm 0.029$ & $0.3852 \pm 0.0029$ & $0.4974 \pm 0.0035$ \\
$2458503.66565834$ &   $16.8097 \pm 0.0086$ & $0.041 \pm 0.007$ & $15.6483$ & $179$ & $0.1081 \pm 0.0010$ & $-4.775 \pm 0.028$ & $0.3916 \pm 0.0021$ & $0.4952 \pm 0.0025$ \\
$2458521.58681966$ &   $16.8295 \pm 0.0087$ & $0.030 \pm 0.007$ & $15.5953$ & $176$ & $0.1123 \pm 0.0010$ & $-4.737 \pm 0.026$ & $0.3975 \pm 0.0021$ & $0.4946 \pm 0.0027$ \\
$2458542.59373287$ &   $16.8271 \pm 0.0090$ & $0.031 \pm 0.007$ & $15.6742$ & $170$ & $0.1095 \pm 0.0010$ & $-4.762 \pm 0.028$ & $0.3817 \pm 0.0022$ & $0.4945 \pm 0.0027$ \\
$2458543.57904173$ &   $16.8330 \pm 0.0083$ & $0.024 \pm 0.007$ & $15.6701$ & $191$ & $0.1076 \pm 0.0009$ & $-4.760 \pm 0.027$ & $0.3774 \pm 0.0020$ & $0.4961 \pm 0.0024$ \\
$2458544.66708564$ &   $16.8440 \pm 0.0089$ & $0.008 \pm 0.007$ & $15.7129$ & $171$ & $0.1067 \pm 0.0010$ & $-4.804 \pm 0.034$ & $0.3864 \pm 0.0022$ & $0.4961 \pm 0.0028$ \\
$2458545.62781982$ &   $16.8214 \pm 0.0101$ & $0.031 \pm 0.008$ & $15.6486$ & $141$ & $0.1071 \pm 0.0013$ & $-4.792 \pm 0.033$ & $0.3856 \pm 0.0028$ & $0.4884 \pm 0.0035$ \\
$2458546.64551855$ &   $16.8438 \pm 0.0106$ & $0.043 \pm 0.008$ & $15.7462$ & $132$ & $0.1109 \pm 0.0014$ & $-4.804 \pm 0.036$ & $0.3896 \pm 0.0029$ & $0.4950 \pm 0.0036$ \\
$2458547.57166286$ &   $16.8337 \pm 0.0101$ & $0.023 \pm 0.008$ & $15.6793$ & $141$ & $0.1092 \pm 0.0013$ & $-4.794 \pm 0.032$ & $0.3938 \pm 0.0028$ & $0.4926 \pm 0.0034$ \\
$2458548.62966483$ &   $16.8329 \pm 0.0087$ & $0.017 \pm 0.007$ & $15.7019$ & $179$ & $0.1068 \pm 0.0010$ & $-4.777 \pm 0.029$ & $0.3909 \pm 0.0022$ & $0.4973 \pm 0.0028$ \\
$2458550.64407369$ &   $16.8320 \pm 0.0085$ & $0.023 \pm 0.007$ & $15.6949$ & $184$ & $0.1087 \pm 0.0010$ & $-4.816 \pm 0.033$ & $0.3901 \pm 0.0021$ & $0.4930 \pm 0.0027$ \\
$2458554.63465672$ &   $16.8501 \pm 0.0103$ & $0.019 \pm 0.008$ & $15.7519$ & $138$ & $0.1140 \pm 0.0014$ & $-4.781 \pm 0.033$ & $0.4019 \pm 0.0028$ & $0.4969 \pm 0.0035$ \\
$2458555.62085265$ &   $16.8363 \pm 0.0087$ & $0.026 \pm 0.007$ & $15.7304$ & $177$ & $0.1089 \pm 0.0010$ & $-4.767 \pm 0.029$ & $0.4028 \pm 0.0022$ & $0.4978 \pm 0.0027$ \\
$2458556.65947799$ &   $16.8486 \pm 0.0111$ & $0.001 \pm 0.009$ & $15.7611$ & $124$ & $0.1107 \pm 0.0015$ & $-4.797 \pm 0.041$ & $0.4070 \pm 0.0032$ & $0.4969 \pm 0.0037$ \\
$2458558.59692547$ &   $16.8513 \pm 0.0113$ & $0.035 \pm 0.009$ & $15.7526$ & $121$ & $0.1134 \pm 0.0016$ & $-4.791 \pm 0.034$ & $0.4061 \pm 0.0032$ & $0.4972 \pm 0.0038$ \\
$2458560.65048512$ &   $16.8578 \pm 0.0097$ & $0.020 \pm 0.008$ & $15.8235$ & $152$ & $0.1092 \pm 0.0013$ & $-4.782 \pm 0.036$ & $0.3930 \pm 0.0027$ & $0.4934 \pm 0.0031$ \\
$2458566.60098112$ &   $16.8672 \pm 0.0084$ & $0.026 \pm 0.007$ & $15.7052$ & $187$ & $0.1078 \pm 0.0009$ & $-4.763 \pm 0.030$ & $0.3895 \pm 0.0021$ & $0.4959 \pm 0.0025$ \\
$2458568.56334852$ &   $16.8670 \pm 0.0087$ & $0.038 \pm 0.007$ & $15.6716$ & $177$ & $0.1098 \pm 0.0010$ & $-4.749 \pm 0.028$ & $0.3948 \pm 0.0022$ & $0.4946 \pm 0.0026$ \\
$2458569.55147254$ &   $16.8569 \pm 0.0090$ & $0.039 \pm 0.007$ & $15.6554$ & $168$ & $0.1100 \pm 0.0011$ & $-4.780 \pm 0.030$ & $0.3890 \pm 0.0023$ & $0.4970 \pm 0.0028$ \\
$2458570.59258685$ &   $16.8796 \pm 0.0106$ & $0.001 \pm 0.008$ & $15.6557$ & $132$ & $0.1094 \pm 0.0014$ & $-4.789 \pm 0.035$ & $0.3898 \pm 0.0030$ & $0.4984 \pm 0.0035$ \\
$2458572.59221071$ &   $16.8804 \pm 0.0099$ & $0.029 \pm 0.008$ & $15.7094$ & $146$ & $0.1083 \pm 0.0013$ & $-4.746 \pm 0.031$ & $0.3868 \pm 0.0027$ & $0.5050 \pm 0.0034$ \\
$2458595.50775173$ &   $16.8567 \pm 0.0137$ & $0.026 \pm 0.010$ & $15.5962$ & $ 93$ & $0.1176 \pm 0.0020$ & $-4.822 \pm 0.046$ & $0.3869 \pm 0.0043$ & $0.4957 \pm 0.0052$ \\
$2458597.49996378$ &   $16.8591 \pm 0.0101$ & $0.008 \pm 0.008$ & $15.6091$ & $143$ & $0.1135 \pm 0.0013$ & $-4.762 \pm 0.031$ & $0.3817 \pm 0.0027$ & $0.4962 \pm 0.0033$ \\
$2458617.50211498$ &   $16.8542 \pm 0.0093$ & $0.022 \pm 0.008$ & $15.7453$ & $161$ & $0.1088 \pm 0.0011$ & $-4.790 \pm 0.044$ & $0.4038 \pm 0.0024$ & $0.4957 \pm 0.0030$ \\
$2458670.93647358$ &   $16.9094 \pm 0.0109$ & $0.025 \pm 0.009$ & $15.6949$ & $126$ & $0.1051 \pm 0.0014$ & $-4.739 \pm 0.054$ & $0.3878 \pm 0.0030$ & $0.4915 \pm 0.0036$ \\
$2458718.87031042$ &   $16.9107 \pm 0.0091$ & $0.012 \pm 0.007$ & $15.6143$ & $165$ & $0.1108 \pm 0.0011$ & $-4.748 \pm 0.028$ & $0.3934 \pm 0.0024$ & $0.4885 \pm 0.0029$ \\
$2458721.92050686$ &   $16.9072 \pm 0.0110$ & $0.016 \pm 0.009$ & $15.6205$ & $124$ & $0.1099 \pm 0.0014$ & $-4.706 \pm 0.028$ & $0.3911 \pm 0.0032$ & $0.4924 \pm 0.0038$ \\
$2458723.89624752$ &   $16.9272 \pm 0.0098$ & $0.015 \pm 0.008$ & $15.6508$ & $147$ & $0.1089 \pm 0.0012$ & $-4.748 \pm 0.029$ & $0.3958 \pm 0.0026$ & $0.4913 \pm 0.0032$ \\
$2458724.92469963$ &   $16.9254 \pm 0.0120$ & $0.024 \pm 0.009$ & $15.6280$ & $111$ & $0.1102 \pm 0.0016$ & $-4.724 \pm 0.031$ & $0.3926 \pm 0.0035$ & $0.4925 \pm 0.0043$ \\
$2458784.85036502$ &   $16.9450 \pm 0.0098$ & $0.013 \pm 0.008$ & $15.5626$ & $147$ & $0.1071 \pm 0.0012$ & $-4.783 \pm 0.030$ & $0.3894 \pm 0.0026$ & $0.4911 \pm 0.0032$ \\
$2458792.78919017$ &   $16.9334 \pm 0.0089$ & $0.019 \pm 0.007$ & $15.6070$ & $171$ & $0.1054 \pm 0.0010$ & $-4.770 \pm 0.028$ & $0.3861 \pm 0.0023$ & $0.4970 \pm 0.0028$ \\
$2458795.78669357$ &   $16.9561 \pm 0.0089$ & $0.023 \pm 0.007$ & $15.5507$ & $170$ & $0.1059 \pm 0.0010$ & $-4.762 \pm 0.028$ & $0.3883 \pm 0.0023$ & $0.4904 \pm 0.0027$ \\
$2458800.72101439$ &   $16.9028 \pm 0.0096$ & $0.019 \pm 0.008$ & $15.5996$ & $150$ & $0.1079 \pm 0.0012$ & $-4.789 \pm 0.031$ & $0.3859 \pm 0.0026$ & $0.4981 \pm 0.0031$ \\
$2458802.84973114$ &   $16.8795 \pm 0.0089$ & $0.025 \pm 0.007$ & $15.5810$ & $171$ & $0.1075 \pm 0.0010$ & $-4.734 \pm 0.026$ & $0.3854 \pm 0.0023$ & $0.5027 \pm 0.0027$ \\
$2458804.71328670$ &   $16.8841 \pm 0.0091$ & $0.038 \pm 0.007$ & $15.6398$ & $165$ & $0.1067 \pm 0.0011$ & $-4.762 \pm 0.028$ & $0.3886 \pm 0.0023$ & $0.4982 \pm 0.0028$ \\
$2458806.69577168$ &   $16.8941 \pm 0.0085$ & $0.019 \pm 0.007$ & $15.6158$ & $182$ & $0.1052 \pm 0.0010$ & $-4.749 \pm 0.027$ & $0.3894 \pm 0.0022$ & $0.5048 \pm 0.0026$ \\
$2458911.59453852$ &   $16.8965 \pm 0.0099$ & $0.028 \pm 0.008$ & $15.7209$ & $147$ & $0.1047 \pm 0.0012$ & $-4.743 \pm 0.028$ & $0.3857 \pm 0.0027$ & $0.4960 \pm 0.0034$ \\
$2458917.63877563$ &   $16.8883 \pm 0.0204$ & $0.021 \pm 0.013$ & $15.7356$ & $ 59$ & $0.1110 \pm 0.0032$ & $-4.732 \pm 0.064$ & $0.3940 \pm 0.0069$ & $0.5077 \pm 0.0082$ \\
\hline
\end{tabular}
}
\end{center}
\end{table*}

\begin{table*}[pht]
\begin{center}
\caption{RV and activity indices obtained from the HARPS spectra.}
\label{tab:harps-data}
\centering
\resizebox{\textwidth}{!}{%
\begin{tabular}{llllllllll}
\hline  \hline
BJD - 2457000 [d] & RV [km/s] & Bisector & FWHM & SNR & H$_\alpha$ & log($R^\prime_{HK}$) & Na~II & He~I \\
\hline
$1464.83460002$	 & $16.8822 \pm 0.0035$ & $0.015 \pm 0.003$ & $14.3788$ & $ 83$ & $0.1084 \pm 0.0008$ & $  -4.771 \pm    0.029$ & $0.4144 \pm 0.0016$ & $0.4977 \pm 0.0019$ \\ 
$1464.82626718$	 & $16.8868 \pm 0.0035$ & $0.017 \pm 0.003$ & $14.3902$ & $ 83$ & $0.1071 \pm 0.0008$ & $  -4.763 \pm    0.029$ & $0.4139 \pm 0.0016$ & $0.4976 \pm 0.0019$ \\
$1465.80299404$	 & $16.9102 \pm 0.0020$ & $0.028 \pm 0.002$ & $14.3445$ & $118$ & $0.1080 \pm 0.0005$ & $  -4.714 \pm    0.025$ & $0.4111 \pm 0.0011$ & $0.4978 \pm 0.0013$ \\
$1465.79547085$	 & $16.9147 \pm 0.0020$ & $0.025 \pm 0.002$ & $14.3513$ & $120$ & $0.1077 \pm 0.0005$ & $  -4.707 \pm    0.025$ & $0.4093 \pm 0.0011$ & $0.4973 \pm 0.0013$ \\
$1466.78878939$	 & $16.9063 \pm 0.0022$ & $0.006 \pm 0.002$ & $14.4190$ & $ 99$ & $0.1079 \pm 0.0006$ & $  -4.701 \pm    0.025$ & $0.4108 \pm 0.0013$ & $0.4956 \pm 0.0015$ \\
$1466.79644105$	 & $16.9112 \pm 0.0020$ & $0.019 \pm 0.002$ & $14.4287$ & $107$ & $0.1074 \pm 0.0006$ & $  -4.675 \pm    0.024$ & $0.4131 \pm 0.0012$ & $0.4949 \pm 0.0014$ \\
$1481.74528823$	 & $16.8882 \pm 0.0020$ & $0.024 \pm 0.002$ & $14.3635$ & $120$ & $0.1061 \pm 0.0006$ & $  -4.727 \pm    0.026$ & $0.4170 \pm 0.0012$ & $0.4979 \pm 0.0013$ \\
$1482.75123537$	 & $16.8830 \pm 0.0020$ & $0.020 \pm 0.002$ & $14.3771$ & $142$ & $0.1062 \pm 0.0004$ & $  -4.722 \pm    0.025$ & $0.4160 \pm 0.0009$ & $0.4982 \pm 0.0011$ \\
$1482.74376858$	 & $16.8829 \pm 0.0020$ & $0.020 \pm 0.002$ & $14.3770$ & $145$ & $0.1057 \pm 0.0004$ & $  -4.740 \pm    0.026$ & $0.4159 \pm 0.0009$ & $0.4965 \pm 0.0011$ \\
$1483.77055220$	 & $16.8847 \pm 0.0020$ & $0.009 \pm 0.002$ & $14.3640$ & $121$ & $0.1066 \pm 0.0005$ & $  -4.736 \pm    0.026$ & $0.4138 \pm 0.0011$ & $0.4978 \pm 0.0013$ \\
$1483.76263412$	 & $16.8846 \pm 0.0020$ & $0.014 \pm 0.002$ & $14.3626$ & $118$ & $0.1067 \pm 0.0005$ & $  -4.747 \pm    0.027$ & $0.4135 \pm 0.0011$ & $0.4975 \pm 0.0013$ \\
$1576.54303679$	 & $16.9129 \pm 0.0029$ & $0.036 \pm 0.002$ & $14.3866$ & $ 90$ & $0.1044 \pm 0.0008$ & $  -4.894 \pm    0.037$ & $0.4165 \pm 0.0015$ & $0.4960 \pm 0.0018$ \\
$1762.82961396$	 & $16.9229 \pm 0.0020$ & $0.014 \pm 0.002$ & $14.4052$ & $113$ & $0.1070 \pm 0.0006$ & $  -4.721 \pm    0.026$ & $0.4196 \pm 0.0012$ & $0.4996 \pm 0.0014$ \\
$1763.79774196$	 & $16.8988 \pm 0.0020$ & $0.030 \pm 0.002$ & $14.3579$ & $114$ & $0.1061 \pm 0.0006$ & $  -4.720 \pm    0.025$ & $0.4225 \pm 0.0012$ & $0.5017 \pm 0.0014$ \\
$1764.83063596$	 & $16.9055 \pm 0.0020$ & $0.025 \pm 0.002$ & $14.3443$ & $139$ & $0.1050 \pm 0.0005$ & $  -4.751 \pm    0.027$ & $0.4237 \pm 0.0010$ & $0.5010 \pm 0.0012$ \\
$1773.87633433$	 & $16.9275 \pm 0.0020$ & $0.014 \pm 0.002$ & $14.3624$ & $109$ & $0.1054 \pm 0.0006$ & $  -4.782 \pm    0.029$ & $0.4233 \pm 0.0013$ & $0.4989 \pm 0.0014$ \\
$1775.86482034$	 & $16.9506 \pm 0.0040$ & $0.004 \pm 0.003$ & $14.3949$ & $ 78$ & $0.1056 \pm 0.0009$ & $  -4.784 \pm    0.030$ & $0.4177 \pm 0.0017$ & $0.4979 \pm 0.0020$ \\
$1777.74548666$	 & $16.9247 \pm 0.0083$ & $0.015 \pm 0.006$ & $14.3560$ & $ 52$ & $0.1058 \pm 0.0015$ & $  -5.023 \pm    0.060$ & $0.4254 \pm 0.0028$ & $0.4988 \pm 0.0031$ \\
$1803.76107908$	 & $16.9097 \pm 0.0020$ & $0.032 \pm 0.002$ & $14.4174$ & $111$ & $0.1064 \pm 0.0006$ & $  -4.757 \pm    0.027$ & $0.4212 \pm 0.0012$ & $0.5007 \pm 0.0014$ \\
$1810.87660156$	 & $16.8759 \pm 0.0020$ & $0.012 \pm 0.002$ & $14.4132$ & $109$ & $0.1054 \pm 0.0006$ & $  -4.771 \pm    0.028$ & $0.4212 \pm 0.0012$ & $0.5010 \pm 0.0014$ \\
$1820.86191239$	 & $16.9179 \pm 0.0020$ & $0.015 \pm 0.002$ & $14.3700$ & $114$ & $0.1052 \pm 0.0006$ & $  -4.766 \pm    0.028$ & $0.4167 \pm 0.0012$ & $0.5006 \pm 0.0013$ \\
$1831.79439764$	 & $16.9347 \pm 0.0020$ & $0.029 \pm 0.002$ & $14.3699$ & $121$ & $0.1043 \pm 0.0006$ & $  -4.761 \pm    0.028$ & $0.4115 \pm 0.0011$ & $0.4982 \pm 0.0012$ \\
$1833.69330017$	 & $16.9464 \pm 0.0030$ & $0.023 \pm 0.002$ & $14.3629$ & $ 88$ & $0.1040 \pm 0.0008$ & $  -4.776 \pm    0.029$ & $0.4138 \pm 0.0015$ & $0.5002 \pm 0.0018$ \\
$1836.70976685$	 & $16.9510 \pm 0.0020$ & $0.022 \pm 0.002$ & $14.3267$ & $107$ & $0.1058 \pm 0.0007$ & $  -4.757 \pm    0.027$ & $0.4183 \pm 0.0013$ & $0.4990 \pm 0.0015$ \\
$1839.73047160$	 & $16.9502 \pm 0.0021$ & $0.021 \pm 0.002$ & $14.3294$ & $100$ & $0.1038 \pm 0.0007$ & $  -4.776 \pm    0.029$ & $0.4232 \pm 0.0014$ & $0.4998 \pm 0.0016$ \\
$1848.70736633$	 & $16.9491 \pm 0.0020$ & $0.042 \pm 0.002$ & $14.3724$ & $152$ & $0.1045 \pm 0.0004$ & $  -4.760 \pm    0.027$ & $0.4180 \pm 0.0009$ & $0.4977 \pm 0.0010$ \\
$1853.82221087$	 & $16.9462 \pm 0.0020$ & $0.018 \pm 0.002$ & $14.3981$ & $106$ & $0.1044 \pm 0.0006$ & $  -4.748 \pm    0.028$ & $0.4055 \pm 0.0012$ & $0.4949 \pm 0.0014$ \\
$1855.62240768$	 & $16.9285 \pm 0.0020$ & $0.011 \pm 0.002$ & $14.3667$ & $113$ & $0.1045 \pm 0.0006$ & $  -4.758 \pm    0.027$ & $0.4148 \pm 0.0012$ & $0.4974 \pm 0.0014$ \\
$1866.53795150$	 & $16.9294 \pm 0.0336$ & $0.043 \pm 0.025$ & $14.2712$ & $ 17$ & $0.0828 \pm 0.0058$ & $\cdots$ & $0.4236 \pm 0.0095$ & $0.4839 \pm 0.0090$ \\
$1866.55739621$	 & $16.9209 \pm 0.0023$ & $0.029 \pm 0.002$ & $14.3386$ & $ 97$ & $0.1031 \pm 0.0007$ & $  -4.759 \pm    0.028$ & $0.4061 \pm 0.0014$ & $0.4969 \pm 0.0016$ \\
$1868.53027483$	 & $16.9340 \pm 0.0020$ & $0.016 \pm 0.002$ & $14.3383$ & $116$ & $0.1048 \pm 0.0006$ & $  -4.761 \pm    0.027$ & $0.4005 \pm 0.0011$ & $0.4917 \pm 0.0013$ \\
$1875.52489656$	 & $16.9384 \pm 0.0020$ & $0.020 \pm 0.002$ & $14.3597$ & $132$ & $0.1041 \pm 0.0005$ & $  -4.752 \pm    0.027$ & $0.4081 \pm 0.0010$ & $0.4947 \pm 0.0011$ \\
$1877.62220328$	 & $16.9245 \pm 0.0049$ & $0.028 \pm 0.004$ & $14.3702$ & $ 71$ & $0.1015 \pm 0.0009$ & $  -4.786 \pm    0.030$ & $0.4067 \pm 0.0017$ & $0.4981 \pm 0.0021$ \\
$1884.61075979$	 & $16.9418 \pm 0.0056$ & $0.030 \pm 0.004$ & $14.3274$ & $ 66$ & $0.1051 \pm 0.0010$ & $  -4.792 \pm    0.031$ & $0.4170 \pm 0.0020$ & $0.4988 \pm 0.0023$ \\
$1886.65333879$	 & $16.9538 \pm 0.0031$ & $0.027 \pm 0.002$ & $14.3517$ & $ 87$ & $0.1047 \pm 0.0007$ & $  -4.824 \pm    0.032$ & $0.4172 \pm 0.0015$ & $0.5010 \pm 0.0017$ \\
$1889.64841759$	 & $16.9595 \pm 0.0033$ & $0.028 \pm 0.002$ & $14.3437$ & $ 85$ & $0.1050 \pm 0.0007$ & $  -4.808 \pm    0.031$ & $0.4221 \pm 0.0015$ & $0.4991 \pm 0.0018$ \\
$1893.71018832$	 & $16.9813 \pm 0.0020$ & $0.020 \pm 0.002$ & $14.3765$ & $104$ & $0.1049 \pm 0.0005$ & $  -4.804 \pm    0.031$ & $0.4128 \pm 0.0012$ & $0.4977 \pm 0.0013$ \\
$1896.53584191$	 & $16.9733 \pm 0.0038$ & $0.024 \pm 0.003$ & $14.3444$ & $ 80$ & $0.1049 \pm 0.0008$ & $  -4.810 \pm    0.031$ & $0.4067 \pm 0.0016$ & $0.4951 \pm 0.0018$ \\
$1901.52855834$	 & $16.9797 \pm 0.0020$ & $0.012 \pm 0.002$ & $14.3447$ & $120$ & $0.1043 \pm 0.0005$ & $  -4.744 \pm    0.027$ & $0.4150 \pm 0.0011$ & $0.4980 \pm 0.0013$ \\
\hline
\end{tabular}
}
\end{center}
\end{table*}

\newpage

\begin{table*}[pht]
\begin{center}
\caption{RV obtained from the CORALIE spectra.}
\label{tab:coralie-data}
\centering
\resizebox{\textwidth}{!}{%
\begin{tabular}{llllrllll}
\hline  \hline
BJD - 2457000 [d] & RV [km/s] & FWHM & Contrast & Bisector & Noise & $\mathrm{SNR_{10}}$ & $\mathrm{SNR_{50}}$ & $\mathrm{SNR_{60}}$ \\
\hline
$1462.613463$ & $16.830 \pm 0.012$ & $14.04027$ & $22.731$ & $-0.01808$ & $0.01151$ & $13.90$ & $43.90$ & $45.80$ \\
$1473.766468$ & $16.861 \pm 0.010$ & $14.00994$ & $22.661$ & $-0.01234$ & $0.00921$ & $19.20$ & $51.10$ & $53.00$ \\
$1489.763017$ & $16.793 \pm 0.017$ & $13.99526$ & $22.967$ & $0.00591$ & $0.01640$ & $9.30$ & $33.10$ & $34.00$ \\
$1490.818784$ & $16.823 \pm 0.015$ & $13.99797$ & $22.900$ & $-0.02211$ & $0.01495$ & $9.80$ & $36.10$ & $37.60$ \\
$1528.633488$ & $16.889 \pm 0.011$ & $14.04580$ & $22.585$ & $-0.03084$ & $0.01032$ & $14.70$ & $48.60$ & $51.10$ \\
$1533.663477$ & $16.844 \pm 0.012$ & $14.02184$ & $22.697$ & $-0.01156$ & $0.01184$ & $14.10$ & $41.90$ & $44.30$ \\
$1537.619293$ & $16.842 \pm 0.011$ & $13.99256$ & $22.683$ & $0.00737$ & $0.01087$ & $15.70$ & $44.40$ & $46.10$ \\
$1544.675520$ & $16.833 \pm 0.012$ & $14.03582$ & $22.648$ & $0.02580$ & $0.01146$ & $13.60$ & $43.90$ & $45.60$ \\
$1562.609684$ & $16.822 \pm 0.015$ & $14.03330$ & $22.664$ & $-0.00991$ & $0.01486$ & $10.30$ & $36.00$ & $37.30$ \\
$1569.594671$ & $16.838 \pm 0.011$ & $14.04995$ & $22.601$ & $-0.01878$ & $0.01042$ & $14.50$ & $47.80$ & $49.30$ \\
\hline
\end{tabular}
}
\end{center}
\end{table*}

\newpage
\startlongtable
\begin{deluxetable}{ll}
\tablecaption{RV obtained from the \textsc{Minerva}-Australis spectra.}
\label{tab:minerva-data}
\tablehead{\colhead{BJD - 2457000 [d] } & \colhead{RV [km/s]} }
\startdata
$1486.154190$ &	$-0.020	\pm 0.016$ \\
$1486.168657$ &	$0.027	\pm 0.014$ \\
$1486.183125$ &	$-0.006	\pm 0.014$ \\
$1494.097685$ &	$0.002	\pm 0.013$ \\
$1494.117512$ &	$0.027	\pm 0.012$ \\
$1495.176447$ &	$0.004	\pm 0.012$ \\
$1495.190914$ &	$-0.004	\pm 0.012$ \\
$1496.060648$ &	$-0.006	\pm 0.013$ \\
$1496.075104$ &	$0.008	\pm 0.013$ \\
$1497.067894$ &	$0.019	\pm 0.011$ \\
$1497.082350$ &	$-0.020	\pm 0.011$ \\
$1498.078553$ &	$0.039	\pm 0.013$ \\
$1498.093021$ &	$-0.006	\pm 0.013$ \\
$1499.073831$ &	$0.018	\pm 0.017$ \\
$1499.088299$ &	$0.032	\pm 0.016$ \\
$1500.058241$ &	$0.005	\pm 0.014$ \\
$1500.072708$ &	$-0.003	\pm 0.012$ \\
$1501.100590$ &	$-0.009	\pm 0.013$ \\
$1501.115046$ &	$0.009	\pm 0.013$ \\
$1501.129514$ &	$-0.009	\pm 0.012$ \\
$1502.071400$ &	$-0.015	\pm 0.016$ \\
$1502.085856$ &	$-0.059	\pm 0.016$ \\
$1502.100324$ &	$0.019	\pm 0.015$ \\
$1503.098032$ &	$-0.021	\pm 0.010$ \\
$1503.119433$ &	$-0.009	\pm 0.011$ \\
$1522.930428$ &	$0.038	\pm 0.016$ \\
$1527.003877$ &	$-0.020	\pm 0.024$ \\
$1529.084062$ &	$-0.005	\pm 0.017$ \\
$1529.095058$ &	$-0.025	\pm 0.019$ \\
$1531.081898$ &	$0.012	\pm 0.016$ \\
$1531.092894$ &	$-0.026	\pm 0.018$ \\
$1533.118935$ &	$-0.087	\pm 0.045$ \\
$1534.047905$ &	$-0.020	\pm 0.046$ \\
$1536.053831$ &	$-0.007	\pm 0.020$ \\
$1536.075231$ &	$-0.028	\pm 0.016$ \\
$1537.015012$ &	$-0.057	\pm 0.025$ \\
$1537.036424$ &	$-0.063	\pm 0.021$ \\
$1537.057824$ &	$-0.052	\pm 0.021$ \\
$1539.062593$ &	$-0.116	\pm 0.020$ \\
$1539.077060$ &	$-0.048	\pm 0.041$ \\
$1551.995208$ &	$-0.059	\pm 0.018$ \\
$1552.017650$ &	$-0.028	\pm 0.020$ \\
$1555.019236$ &	$-0.001	\pm 0.014$ \\
$1555.040637$ &	$-0.036	\pm 0.015$ \\
$1561.005451$ &	$-0.058	\pm 0.018$ \\
$1562.024977$ &	$-0.101	\pm 0.014$ \\
$1563.003519$ &	$-0.013	\pm 0.041$ \\
$1563.024919$ &	$-0.038	\pm 0.018$ \\
$1566.002882$ &	$-0.004	\pm 0.015$ \\
$1566.024294$ &	$0.002	\pm 0.015$ \\
$1566.953866$ &	$0.003	\pm 0.013$ \\
$1566.975278$ &	$-0.012	\pm 0.014$ \\
$1574.970301$ &	$-0.016	\pm 0.020$ \\
$1574.991701$ &	$-0.012	\pm 0.017$ \\
$1576.979456$ &	$0.003	\pm 0.017$ \\
$1577.967188$ &	$0.016	\pm 0.015$ \\
$1578.968461$ &	$-0.023	\pm 0.016$ \\
$1586.017812$ &	$-0.001	\pm 0.015$ \\
$1586.981354$ &	$-0.034	\pm 0.014$ \\
$1587.002766$ &	$0.053	\pm 0.016$ \\
$1587.024167$ &	$0.036	\pm 0.018$ \\
$1588.962361$ &	$-0.041	\pm 0.015$ \\
\enddata
\end{deluxetable}

\bibliography{sample63}{}
\bibliographystyle{aasjournal}



\end{document}